\newif\ifFINAL
\FINALtrue
\newif\ifUSEIEEE
\USEIEEEtrue
\ifUSEIEEE
  
  \documentclass[10pt,journal,compsoc]{IEEEtran}
  \newcommand{\ParStart}[2]{\IEEEPARstart{#1}{#2}}
  \newcommand{\citep}[1]{\cite{#1}}
  \def\sep{, }
\else
  
  \ifFINAL
    \documentclass[number,preprint]{elsarticle}
  \else
    \documentclass[number,review]{elsarticle}
  \fi
  \newcommand{\ParStart}[2]{\noindent{#1}{#2}}

  \journal{Journal of Parallel and Distributed Computing}

  \usepackage{wrapfig}
  \ifFINAL\else
    \usepackage{setspace}
  \fi
\fi

\usepackage{cite}
\usepackage{amsmath,amssymb,amsfonts}
\usepackage{graphicx}
\usepackage[dvipsnames]{xcolor}
\usepackage{listings}
\usepackage{import}
\usepackage{url}
\ifUSEIEEE\else
  \usepackage{pdfpages}
\fi
\usepackage{hyperref}
\hypersetup{
  colorlinks,
  linkcolor=blue
}
\usepackage[capitalize,nameinlink,noabbrev]{cleveref}

\definecolor{navyblue}{rgb}{0.0, 0.0, 0.5}
\definecolor{indiagreen}{rgb}{0.07, 0.53, 0.03}
\newcommand{\CodeSymbol}[1]{\textcolor{RedOrange}{#1}}

\lstdefinestyle{customcpp}{
  belowcaptionskip=1\baselineskip,
  breaklines=true,
  frame=single,
  language=C++,
  showstringspaces=false,
  basicstyle=\small\ttfamily,
  directivestyle=\color{navyblue},
  keywordstyle=\color{indiagreen},
  keywordstyle={[2]{\color{RedViolet}\bfseries}},
  keywordstyle={[3]{\color{blue}}},
  keywordstyle={[4]{\color{orange}}},
  commentstyle={\itshape\color{red}},
  stringstyle={\color{OliveGreen}},
  morekeywords={
    PetscReal, PetscDeviceContext, PetscDevice, ManagedScalar, 
    ManagedReal, PetscObjectId, PetscMemoryAccessMode, PetscObject, 
    Vec, PetscErrorCode, PetscScalar, PetscPointerAttributes, PetscInt,
    PetscDeviceContextJoinMode, PetscBool, ManagedMemory,
    EvaluatedExpression, PetscStreamType, vector, PetscLogDouble,
    cudaStream_t, Expression, ConstantExpression, queue, SomeObject,
    PDC, KSP, device_event, Mat, ExecutableExpression
  },
  morekeywords=[2]{
    NORM_2, PETSC_MEMORY_ACCESS_WRITE, PETSC_MEMORY_ACCESS_READ,
    PETSC_MEMORY_ACCESS_READ_WRITE, PETSC_MEMTYPE_HIP,
    PETSC_UNKNOWN_MEMORY_ID, PETSC_STREAM_GLOBAL_BLOCKING, 
    PETSC_STREAM_DEFAULT_BLOCKING, PETSC_STREAM_GLOBAL_NONBLOCKING,
    nullptr
  },
  morekeywords=[3]{
    VecNorm, VecScale, LibrarySetStream, MatMult, PetscObjectGetId,
    PetscDeviceContextMarkIntentFromIDBegin,
    PetscDeviceContextMarkIntentFromIDEnd,
    PetscDeviceContextMarkIntentBegin, 
    PetscDeviceContextMarkIntentEnd,
    PetscDeviceContextGetStreamHandle, MyVecScale, 
    PetscDeviceRegisterMemory, VecGetArray, VecGetArrayRead,
    MatDenseGetArrayWrite, PetscDeviceContextFork,
    PetscDeviceContextJoin, PetscDeviceContextSynchronize,
    PetscDeviceContextQueryIdle, PetscDeviceContextWaitForContext,
    VecNormAsync, VecScaleAsync, Eval, MultiEval, cos, max, min, abs,
    Execute, PetscDeviceContextCreate, PetscDeviceContextSetStreamType,
    sin, VecAXPYAsync, PetscDeviceGetPointerAttributes, KSPSolve,
    ResetOperators, PetscTime, cudaDeviceSynchronize, emplace_back,
    PreSolve, PostSolve, submit, MPI_Allreduce, scale_kernel, Foo,
    Bar, Baz, write, read, PetscSin, VecAYPXAsync, VecCopyAsync,
    KSP_MatMultAsync, VecDotAsync, KSP_PCApplyAsync, converged, front,
    VecAYPX, KSP_MatMult, VecRestoreArrayRead, VecGetLocalSize,
    KSP_PCApply, cudaEventSynchronize, cublasGetStream, 
    cublasSetStream, End, ApplyPreconditioner,
    user_converged_callback, cublasSetPointerMode,
    cudaStreamSetAttribute
  },
  emphstyle=\CodeSymbol,
  emph={
    const, static, inline, constexpr, alignof, decltype, sizeof, for, 
    enum, if, break
  },
  literate={
    {>}{{\CodeSymbol{>}}}1
    {<}{{\CodeSymbol{<}}}1
    {=}{{\CodeSymbol{=}}}1
    {&}{{\CodeSymbol{\&}}}1
    {*}{{\CodeSymbol{*}}}1
    {/\ }{{\CodeSymbol{/\ }}}1
    {-}{{\CodeSymbol{-}}}1
    {+}{{\CodeSymbol{+}}}1},
}

\makeatletter
\def\lst@makecaption{%
  \def\@captype{table}%
  \@makecaption%
}%
\makeatother

\def\PaperAbstract{Leveraging Graphics Processing Units (GPUs) to accelerate scientific software has proven to be highly successful, but in order to extract more performance, GPU programmers must overcome the high latency costs associated with their use. One method of reducing or hiding this latency cost is to use asynchronous \emph{streams} to issue commands to the GPU. While performant, the streams model is an invasive abstraction, and has therefore proven difficult to integrate into general-purpose libraries. In this work, we enumerate the difficulties specific to library authors in adopting streams, and present recent work on addressing them. Finally, we present a unified asynchronous programming model for use in the Portable, Extensible, Toolkit for Scientific Computation (PETSc) to overcome these challenges. The new model shows broad performance benefits while remaining ergonomic to the user.}
\def\PaperKeywords{GPU\sep Streams\sep PETSc\sep Asynchronous}

\title{Safe, Seamless, And Scalable Integration Of Asynchronous GPU Streams In PETSc}

\begin{document}

\includepdf[pages={1}]{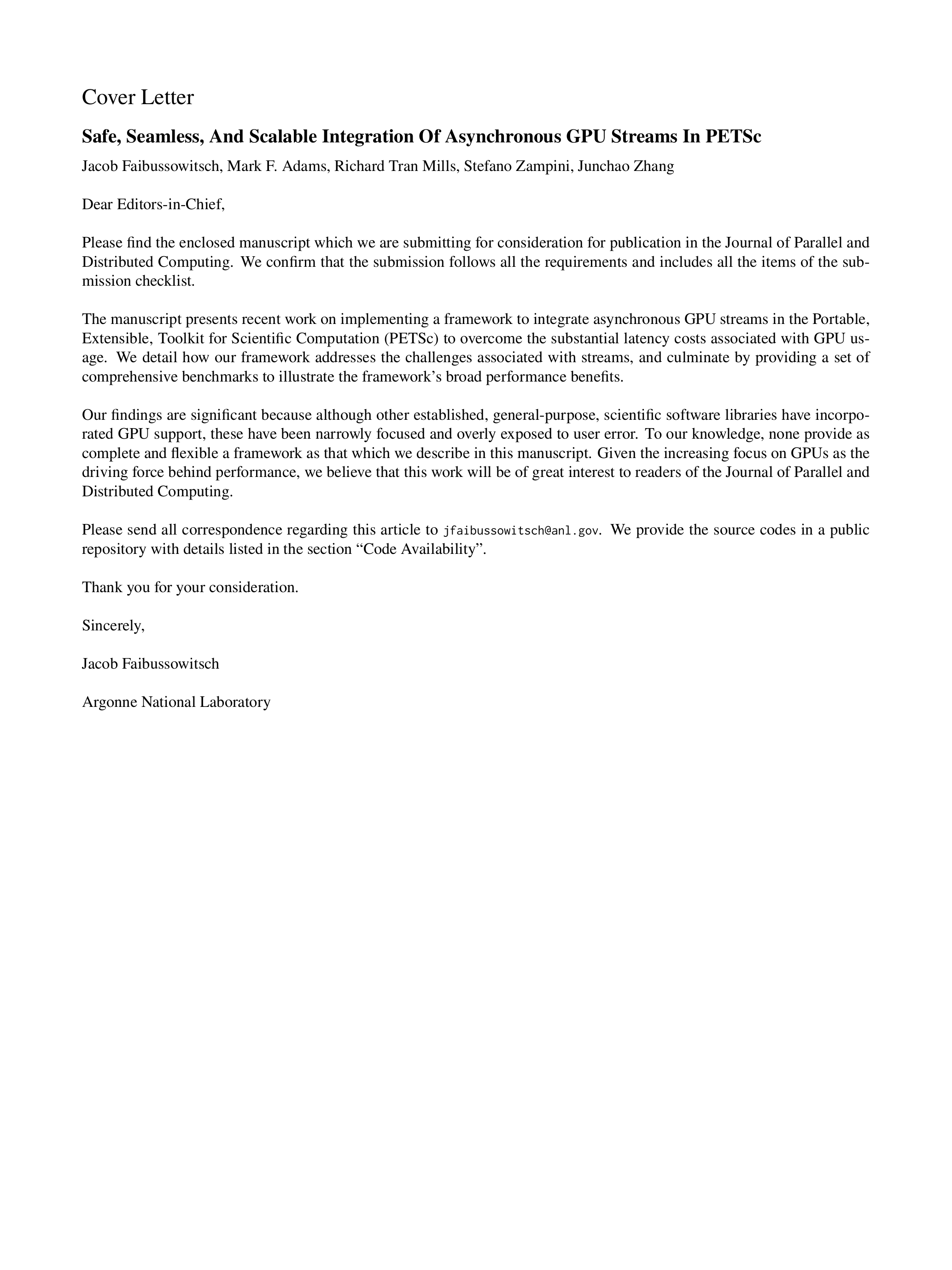}

\begin{frontmatter}

\author[1]{Jacob~Faibussowitsch\corref{cor1}}
\ead{jfaibussowitsch@anl.gov}
\author[2]{Mark~F.~Adams}
\ead{mfadams@lbl.gov}
\author[1]{Richard~Tran~Mills}
\ead{rtmills@anl.gov}
\author[3]{Stefano~Zampini}
\ead{stefano.zampini@kaust.edu.sa}
\author[1]{Junchao~Zhang}
\ead{jczhang@anl.gov}

\cortext[cor1]{Corresponding author}

\affiliation[1]{
organization={Argonne~National~Laboratory},
addressline={9700~S.~Cass~Avenue},
postcode={60439},
city={Lemont},
state={IL},
country={U.S.A.}}

\affiliation[2]{
organization={Lawrence~Berkely~National~Laboratory},
addressline={1~Cyclotron~Road},
postcode={94720},
city={Berkeley},
state={CA},
country={U.S.A.}}

\affiliation[3]{
organization={Extreme~Computing~Research~Center,\\King~Abdullah~University~of~Science~and~Technology},
postcode={23955-6900},
city={Thuwal},
country={Saudi~Arabia}}

\ifFINAL\else
  \newpageafter{author}
\fi

\begin{abstract}
  \PaperAbstract    
\end{abstract}

\begin{keyword}
  \PaperKeywords
\end{keyword}

\end{frontmatter}

\ifFINAL\else
  \pagebreak
\fi

\section{Introduction}
\label{sec:intro}
\ParStart{H}{igh}-Performance Computing (HPC) has been transformed by the introduction of the \textit{Graphics Processing Unit} (GPU) whose massively parallel architecture is able to accelerate many traditionally CPU-based algorithms. While GPUs are a powerful tool to improve on-node performance, they have nevertheless proven difficult to integrate into established HPC libraries.

GPU architectures present several challenges to scientific software developers; see for example~\citep{MILLS2021}. Chief among these is that GPU designs favor high computational throughput over low latency~\citep{Latency}. Developers must design algorithms to avoid, or mask, significant latency associated with kernel launches, transfers between host and device memory, and barrier synchronization just to name a few. Some of these latencies can be effectively amortized by simply ensuring that GPU kernels are launched with a sufficiently large quantity of work. But this is not always practicable.

GPU vendors have sought to offset these latency costs by introducing the concept of \textit{streams}, which act like queues to which operations can be submitted. This paper develops a light-weight GPU-oriented programming model that reduces latency costs by efficiently and transparently leveraging streams in the context of the \emph{Portable, Extensible Toolkit for Scientific Computation} (PETSc)\citep{petsc-user-ref}, which is a general purpose library designed to solve large-scale linear and nonlinear algebraic systems and optimization problems.

Streams share many similarities with POSIX threads~\citep{POSIX}: all streams have access to the same memory space, inter-stream communication is done using semaphore-like ``events'', and programmers can define data dependencies by serializing multiple streams via events. Naturally, streams and threading also share the same concept of sequential consistency. Work submitted to the same stream is guaranteed to execute sequentially, but work submitted to different streams may execute concurrently.

The upshot is that while streams do effectively hide the latency, it is no longer deterministic when enqueued work completes. There is no assumption of forward progress until the programmer manually re-establishes coherence by synchronizing a stream, and since synchronizing needs to contend with possibly multiple interdependent streams, it carries additional costs and is generally avoided wherever possible.

Streams have posed a serious integration challenge for library authors, which, unlike application codes, are beholden to their users. Such authors must establish a --- possibly future-proof --- Application Programming Interface (API), and make clear any assumptions about context (e.g., device or host code, thread-safe, or not), whereas monolithic application codes can modify an internal API at will. Unless the library is expressly GPU- or parallelism-oriented, implementing a stream-like interface often represents a major departure from user-held assumptions about a library's interface. These assumptions can be categorized as three major problems: the \emph{asynchronous} problem, the \emph{concurrent} problem, and the \emph{scalar} problem.

This work is organized as follows: \cref{sec:problems} describes each problem in detail, and provides examples of how contemporary approaches by others are deficient or sub-optimal. \cref{sec:design} begins by developing concrete solutions to these problems and culminates in presenting a unified framework to solve the stream-integration issue writ-large. Finally, we culminate with a comprehensive set of performance benchmarks of the framework in \cref{sec:results} to demonstrate the effectiveness of our proposal.
\section{Problems And Related Work}
\label{sec:problems}
\subsection{The Asynchronous Problem}
\label{sec:problems:async_problem}
\noindent
The asynchronous nature of the streams model breaks the assumption of \emph{atomicity}. That is, a caller of a routine assumes the callee is done executing after it returns and the result it produced is immediately consumable. By extension, they also expect any side effects as a result of this call to be immediately visible.

For example, suppose a user calls a routine to calculate and return the norm of a vector. Logically, the user would expect the resultant norm to be valid immediately after the routine returns and might use it as input in subsequent operations. However, if the underlying implementation of the routine is done on the GPU using streams, it may not even have launched yet, let alone be ready. This begs the question, how should a library communicate this with users? Historically, libraries have been left with the following choices:
\begin{itemize}
    \item \textbf{Use streams internally but synchronize after every GPU operation}. This is both safe and well-encapsulated: the user can trust that results are immediately available (just like on the CPU) but it is inefficient. Each call, no matter how compute-intensive, must pay the full cost of establishing coherence, and synchronizing stifles any opportunity for hiding latency. 
    
    This approach is commonly taken by ostensibly GPU-aware Message Passing Interface (MPI) implementations, which may directly take device buffers as MPI routine arguments. For example, to implement \lstinline{MPI_Allreduce()} with device buffers, the implementation must synchronize its internal streams before returning control to the user in order to ensure the result is valid.

    \item \textbf{Return event objects that users can synchronize themselves}. This allows the user more fine-grained control as they may defer synchronization. For example, SYCL~\citep{SYCL} and Thrust~\citep{THRUST} support this approach. Calling the \lstinline{submit()} method on a SYCL \lstinline{queue} object will return a SYCL event for users to wait on. Thrust exposes a selection of algorithms under the \lstinline{thrust::async} namespace which all return a \lstinline{thrust::device_event}.

    While more performant, this approach is nevertheless unsavory. The caller now needs to store the returned events, or pass them further up the call chain, leaking implementation details. It is also difficult to compose such an approach with multiple cooperating libraries, which might all have their own flavor of events. In general, this strategy makes generic code harder to write as the myriad of events pollutes the business logic.

    For example, the aforementioned \lstinline{thrust::device_event}s call \lstinline{cudaEventSynchronize()} in their destructor, forcing the caller to store the events to avoid this. Since a library cannot know \emph{a priori} when the user actually wants to synchronize, they have no choice but to either return the events, destroy them immediately, or store them indefinitely.

    \item \textbf{Directly expose the stream the library is using}. This approach has the benefit of maximum flexibility and is commonly taken by the various GPU-vendor libraries. For example, NVIDIA's cuBLAS library exposes \lstinline{cublasGetStream()} and \lstinline{cublasSetStream()}.
    
    On the one hand, the user is free to do whatever they wish with the stream. But on the other hand, \textit{the user is free to do whatever they wish with the stream}. The user may elect to destroy the stream out from underneath the library or change some behavior of the stream upon which the library implicitly relies on (for example, users are free to call \lstinline{cudaStreamSetAttribute()}
to change attributes of the CUDA stream a library is using). 
    Or they may misuse the stream, causing some large performance degradation. On balance, relinquishing control of the underlying stream is a \textit{Pandora's box}.
\end{itemize}
In summary, these choices boil down to two competing philosophies. Either the library needs to care about synchronization, or the user does. Faced with the choice, many library authors have opted for the former as it is both easier to implement and maintain.
\subsection{The Concurrent Execution Problem}
\label{sec:problems:concurrent_problem}
\noindent
Concurrent execution breaks the assumption of \emph{linearity}---that routines are executed in the order in which they appear in a program. For ``normal'' programs, and programs using only a single stream, this pattern is self-evident. If, however, multiple asynchronous streams were used, there would be no telling in what order they would execute. This non-linearity has profound implications.

Streams allow users to introduce race conditions, deadlocks, and other hazards with reckless abandon. While manageable enough within a confined space (such as a low-level function), manually tracking stream dependencies becomes intractable at the library scale. Library authors must also contend with user callbacks, which demand some degree of coherence. There is also the question of how such a model might interact with communication. Even if the user properly delegates a library's stream -- as noted in \cref{sec:problems:async_problem} -- the current MPI specification lacks the ability to reason about the stream. Some implementations (for example PETSc~\citep{PetscSF2022}) have therefore been forced to roll their own asynchronous communication primitives to bridge this gap.

A common approach, then, for stream-aware libraries is to expose some sort of \lstinline{LibrarySetStream()} routine, which allows them to claim multi-stream support. But as alluded to in  \cref{sec:problems:async_problem}, this approach possesses limitations:
\begin{itemize}
    \item \textbf{It introduces global state}. This limits the ability of the library to reason about, or make general optimizations. For example, it may restrict the reuse of scratch buffers since the library cannot guarantee that a previous stream is not still using it.
    \item \textbf{It makes inter-stream dependencies harder to track}.  Conceivably, the current stream may be swapped out at any point; there is an implicit assumption that every routine plays nicely and restores the previous stream after it is done. But this honor system is completely unenforceable programmatically; it obscures the provenance of the stream, making reasoning about it difficult for the code developer and reader.
    \item \textbf{It puts all responsibility on the user}. On the one hand, the caller is best equipped to apply optimizations, and so it seems natural that they make stream serialization decisions for the library. But as before, this causes implementation details to bubble up, making the code less generic and limiting composability. It also removes any possibility of library-assisted error checking, causing subtle bugs to go unnoticed by the user.
\end{itemize}
As before, the choice appears to be between simplicity and maintainability versus maximizing performance.
\subsection{The Scalar Problem}
\label{sec:problems:scalar_problem}
\noindent
This problem differs from those described in \cref{sec:problems:async_problem,sec:problems:concurrent_problem} insofar that is not a subversion of an assumption, rather, it is a purely \emph{logistical} issue. Many algorithms require a small collection of scalar variables. These may be needed to mutate or scale larger objects, or, in linear solvers, for example, to determine convergence of the algorithm. Often they are produced by some routine, trivially modified or inspected, and then passed directly on to a consumer. Put succinctly, the scalar problem is ``given the explicitly massively parallel architecture of the GPU, how can I efficiently pass around single values and do simple arithmetic with them in a generic way?''. To illustrate this, consider the code below:
\begin{lstlisting}[
  caption=Example Implementation of Normalizing a Vector,
  label=code:vec_normalize
]
PetscReal norm, alpha;

VecNorm(v, NORM_2, &norm);
alpha = 1.0 / norm;
VecScale(v, alpha);
\end{lstlisting}
While seemingly reasonable, this process is sub-optimal for device-accelerated libraries. Although the library might efficiently compute the value of \lstinline{norm} on the GPU, it must nevertheless copy the result back to the CPU and synchronize before returning. The same process holds in reverse for \lstinline{VecScale()}. Ideally, the reciprocal is computed on the GPU, meaning the value of \lstinline{norm} never has to leave the device. If the scalar stays on the device, this also has the added benefit of no longer requiring synchronization.

One could argue that since \lstinline{norm} is passed by reference, \lstinline{VecNorm()} could do different things depending on whether the pointer is CPU or GPU memory --- for example, elide copying and synchronizing in the latter case. This approach is taken by cuBLAS, whereby the user may declare (via \lstinline{cublasSetPointerMode()}~\citep{CUBLAS_POINTER_MODE}) that all scalar pointer parameters passed to subsequent cuBLAS functions are either on host or on device. In the case where the pointer is device memory, cuBLAS will (assuming the user has remembered to also set a stream via \lstinline{cublasSetStream()}) launch its kernels asynchronously. 

However, this mechanism provides no safety in case of mistakenly passing the wrong pointer type (and the user must not forget to reset the pointer mode for the next routine). The logical next step would then be for cuBLAS try and determine the pointer's memory type at runtime, but this is equally unsatisfactory. Not only is this check expensive (and unnecessary in hindsight for CPU-only data on a GPU-enabled build\footnote{To make matters worse, the check is also completely unavoidable in these circumstances. Since the raw pointer carries no information about its memory space, the library must assume that \emph{all} user pointers potentially live on the GPU and make this check \emph{everywhere}}), but it misses the forest for the trees. The user would still need to find a way to perform the reciprocal on either the host or device in a \emph{generic} manner. \emph{And then} ensure that \lstinline{VecScale()} is equally well equipped to handle either case gracefully.

Another possible solution consists of providing an implementation for the entire \emph{norm-and-scale} operation; while this sounds reasonable in the simple context considered here, it is not scalable at the library level, where \emph{norm-and-scale} is one of many compute patterns used. In short, it is not sufficient to just consider \emph{where} the memory lives, one must also consider \emph{how} it is used.
\section{Design}
\label{sec:design}
\noindent
To overcome these challenges, this work introduces a novel approach by separating concerns. This is summarized in the following core design philosophies: the library needs to care about serialization, the user needs to care about the parallelism, but \emph{nobody} needs to care about synchronization. The result is a powerful, minimal, and transparent interface.

In \cref{sec:design:dctx,sec:design:managed} we present a pair of new library objects to implement these philosophies. We begin by outlining a set of low-level building blocks used to realize each object and culminate by showing how they concretely addresses the problems outlined in \crefrange{sec:problems:async_problem}{sec:problems:scalar_problem}. Finally, in \cref{sec:design:together}, we present a grand unification of the framework and showcase its use as part of a typical user's workflow.

\subsection{\texorpdfstring{\lstinline{PetscDeviceContext}}{PetscDeviceContext}}
\label{sec:design:dctx}
\noindent
To address the asynchronous and concurrent execution problems we introduce a new library object, \lstinline{PetscDeviceContext}. Conceptually, it is similar to a stream, but in practice it has many other functions and responsibilities. In particular, it serves as a gatekeeper to the various vendor library handles, ensuring that these are always properly managed with the correct stream and device configuration set. It also serves to paper over any differences in vendor implementations (e.g. cuBLAS versus rocBLAS, or cuSPARSE versus rocSPARSE, etc.) so that application code sees a consistent, homogeneous, interface.

To that end, \lstinline{PetscDeviceContext} is comprised of two basic building blocks: the device over which it presides, and the stream that it manages. The stream's behavior is customizable -- manifestly changing the behavior of the \lstinline{PetscDeviceContext} -- by allowing the user to select from one of several stream-types. Among these, ``default blocking'', and ``globally blocking'' are the most common and are described forthwith.

Default blocking streams are both asynchronous with respect to the host, and all other streams. They are analogous to user-created CUDA or HIP streams. Globally blocking streams on the other hand are considered fully host-synchronous. Any work enqueued on the stream will wait for all prior work on all default and globally blocking streams to finish before it starts. These are analogous to the CUDA or HIP ``NULL'' stream~\citep{NULL_STREAM}, with one crucial difference. In addition to waiting for all streams, any operation launched on them is also guaranteed to have completed by the time the launching routine returns.



\subsubsection{Solving The Asynchronous Problem}
\label{sec:design:dctx:async_problem}
\noindent
\lstinline{PetscDeviceContext} tackles the asynchronous problem by providing a mechanism through which the user can \emph{reason} about atomicity. This distinction is subtle. \lstinline{PetscDeviceContext} does not solve the problem outright (in fact, additional tools described hereafter are required to do so), but what functionality \lstinline{PetscDeviceContext} does provide is so foundational to the solution of the asynchronous problem that it cannot be considered solved without it.

The asynchronous problem -- at its core -- boils down to knowing when it is safe to access results. In other words, an asynchronous producer routine must have some way of \emph{communicating} to the consumer when it is safe to inspect the supposed result. To that end, \lstinline{PetscDeviceContext} exposes a safer and streamlined set of functionality over the low-level control provided by raw vendor streams. This is encapsulated by the core stream-control APIs\footnote{These routines all return an \lstinline{enum PetscErrorCode}. As this is the case with all PETSc routines, we have omitted the return type for brevity. The reader can assume that any routine henceforth discussed which does not list a return type, does in fact return \lstinline{PetscErrorCode}.} shown in \cref{code:dctx_core_stream_api}.
\begin{lstlisting}[
  caption=Core PetscDeviceContext Stream Control APIs,
  label=code:dctx_core_stream_api
]
PetscDeviceContextWaitForContext(PetscDeviceContext waiter, PetscDeviceContext waitee);
PetscDeviceContextQueryIdle(PetscDeviceContext dctx, PetscBool *is_idle);
PetscDeviceContextSynchronize(PetscDeviceContext dctx);
\end{lstlisting}
These routines boil down most of the complexity inherent with streams into three simple operations. Make one stream wait for the other, ask and answer the question ``is this stream idle?'', and, finally, synchronize the host with a stream. Conspicuous by omission is any form of ``event'' object or API. This is deliberate. By exposing only high-level stream control, the user can declaratively specify any stream dependencies and need not muddy application code with intermediate objects.

In full compliance with PETSc philosophy, we can have many different backends, including user-generated ones, e.g., a host-only multi-thread backend. So long as the programming model supports the three basic operations detailed above, it may be seamlessly integrated into \lstinline{PetscDeviceContext}.

\subsubsection{Solving The Concurrent Execution Problem}
\label{sec:design:dctx:concurrent_problem}
\noindent
The most important function of \lstinline{PetscDeviceContext} (and that which directly solves the concurrent execution problem), however, is the tracking of inter-stream \textit{dependencies}. This is achieved via a pair of routines described in \cref{code:mark_api}.
\begin{lstlisting}[
  caption=API For Marking Stream Dependencies,
  label=code:mark_api
]
PetscDeviceContextMarkIntentFromIDBegin(PetscDeviceContext dctx, PetscObjectId id, PetscMemoryAccessMode mode, const char description[]);
PetscDeviceContextMarkIntentFromIDEnd(PetscDeviceContext dctx, PetscObjectId id, PetscMemoryAccessMode mode, const char description[]);
\end{lstlisting}
Here \lstinline{id} uniquely identifies a particular object to be marked, \lstinline{mode} describes the memory access pattern (read, write, read-write), and \lstinline{description} provides a descriptive message for debugging purposes. For convenience, a pair of routines \lstinline{PetscDeviceContextMarkIntentBegin/End()} are also provided which take a \lstinline{PetscObject} in place of \lstinline{PetscObjectId}. These routines must enclose the ``critical'' sections where GPU work is launched. An example using CUDA~\citep{CUDA} to launch a scaling kernel is shown in \cref{code:mark_api_example}.
\begin{lstlisting}[
  caption=Example Use of Marking API,
  label=code:mark_api_example
]
PetscErrorCode MyVecScale(Vec v, PetscScalar alpha, PetscDeviceContext dctx)
{
  cudaStream_t *stream;

  PetscDeviceContextMarkIntentBegin(dctx, (PetscObject)v, PETSC_MEMORY_ACCESS_READ_WRITE, "MyVecScale");
  PetscDeviceContextGetStreamHandle(dctx, &stream);
  scale_kernel<<<..., *stream>>>(v->gpu_array, alpha);
  PetscDeviceContextMarkIntentEnd(dctx, (PetscObject)v, PETSC_MEMORY_ACCESS_READ_WRITE, "MyVecScale");
}
\end{lstlisting}
\cref{code:mark_api_example} shows how marking is done on the scale of entire objects, but the marking system can also be applied at the granularity of individual buffers, which is done by first generating or retrieving a unique id for that particular memory region via \lstinline{PetscDeviceRegisterMemory()} (detailed in \cref{code:register_mem}).
\begin{lstlisting}[
  caption=Memory Registration API,
  label=code:register_mem
]
PetscDeviceRegisterMemory(const void *ptr, PetscPointerAttributes *attr);
\end{lstlisting}
Here \lstinline{ptr} points to the start of the desired memory region, and \lstinline{attr} contains the necessary information to describe the memory. Once the memory is successfully registered, the \lstinline{id} field of \lstinline{attr} will be updated with the assigned id. An example of this is shown in \cref{code:register_mem_example}.
\begin{lstlisting}[
  caption=Memory Registration API Example,
  label=code:register_mem_example
]
PetscPointerAttributes attr;

// Memory type of the pointer, in this case
// ptr originates from a call to hipMalloc()
attr.mtype = PETSC_MEMTYPE_HIP;
// As the memory is unregistered, the id of 
// the memory region is not yet known
attr.id = PETSC_UNKNOWN_MEMORY_ID;
// The total size (in bytes) of the region
attr.size = n * sizeof(*ptr);
// The memory alignment (in bytes) of the
// pointer
attr.align = alignof(decltype(*ptr));

PetscDeviceRegisterMemory(ptr, &attr);
// If registration was successful, attr.id now 
// holds the assigned PetscObjectId of the 
// region
PetscDeviceContextMarkIntentFromIDBegin(..., attr.id, ...);
\end{lstlisting}
Alternatively, if the user knows that a particular memory region was previously registered, they may use \lstinline{PetscDeviceGetPointerAttributes()} to directly retrieve the associated \lstinline{PetscPointerAttributes}.


Once an object or memory region has been marked, subsequent accesses to it are ordered according to its \emph{data dependence}. Operations using the same \lstinline{PetscDeviceContext} are ordered as they appear in the source code. That is
\begin{lstlisting}[
  caption=Sequential Ordering Example,
  label=code:same_dctx
]
Foo(dctx);
Bar(dctx);
Baz(dctx);
\end{lstlisting}
are executed as \lstinline{Foo()}, then \lstinline{Bar()}, then \lstinline{Baz()}, regardless of what \lstinline{Foo()}, \lstinline{Bar()}, or \lstinline{Baz()} do. On the other hand, operations using separate \lstinline{PetscDeviceContext}s which access a common object or memory region are strongly write-ordered. That is, the following operations:
\begin{itemize}
    \item write-write
    \item write-read
    \item read-write
\end{itemize}
Are sequenced by the order in which they are executed. In other words:
\begin{lstlisting}[
  caption=Example of Strong Write-Ordering,
  label=code:write_order
]
SomeObject x;
// Both dctx_a and dctx_b perform seemingly conflicting 
// "write" operations on x. But since x.write(dctx_a) 
// appears before x.write(dctx_b) in the source code it 
// is executed before the other.
x.write(dctx_a);
x.write(dctx_b);

// Here the read operation comes after the write, which
// would otherwise conflict. Since the read comes after 
// the write in the source code, it will wait for the write
// to complete before starting.
x.write(dctx_a);
x.read(dctx_b);

// Finally, a write-after-read conflict. As above, the
// apparent conflict is resolved by source code position.
// The write will wait for the read to complete.
x.read(dctx_a);
x.write(dctx_b);
\end{lstlisting}
Note the absence of read-read. There is no prescribed ordering between consecutive read accesses to the same object, regardless of which \lstinline{PetscDeviceContext} is used to do so. Formally: 

\textit{Given an operation A-B (e.g. A = write, B = read) where A,B $\neq$ read, on an object or memory region M such that A “happens-before” B, where A uses \lstinline{PetscDeviceContext} X and B uses \lstinline{PetscDeviceContext} Y (and Y $\neq$ X), then B shall not begin before A completes. This implies that any side-effects resulting from A are also observed by B.}
\subsection{\texorpdfstring{\lstinline{Petsc::ManagedMemory}}{Petsc::ManagedMemory}}
\label{sec:design:managed}
\noindent
To address the final problem -- the scalar problem -- we introduce the \lstinline{ManagedMemory} object, which models a future\footnote{\emph{futures} refer to constructs or objects which act as a proxy for some not-yet-materialized result. These allow the programmer to symbolically represent the result of some asynchronous computation without needing to wait for its concrete value to be produced.} for an opaque, dual array of values. It complements a mirrored host/device array functionality with the ability to symbolically represent expressions and convert them into the corresponding CPU or GPU kernels by way of \emph{expression templates}.

\subsubsection{Expression Templates}
\label{sec:design:managed:expr_templates}
\noindent
As is the case with many expression template implementations, the various arithmetic operators for \lstinline{ManagedMemory} are overloaded to return proxy objects, whose type encodes the action of the operator, and which expose some other function (usually \lstinline{operator[]()}) which performs the action itself. These proxy objects are themselves given the same operator overloading treatment, allowing expressions to recursively combine as needed. In our case, the base proxy object is the template \lstinline{Expression} class. 

Furthermore, special overloads of common mathematical operators (\lstinline{min()}, \lstinline{max()}, \lstinline{abs()}, \lstinline{sin()}, etc.) are also provided to enable wrapping their function in a similar proxy object.  Constants are also supported, by providing overloads that wrap them in a special form of \lstinline{Expression}, the \lstinline{ConstantExpression}. The end result is a system capable of symbolically representing almost any arbitrary expression simply by inspecting the constituent parts at compile time.

The final step is to ``evaluate'' the proxy objects, i.e. to collapse their symbolic representation into a concrete CPU or GPU kernel. This is done via the \lstinline{Eval()} helper routine, which takes an \lstinline{Expression} object and returns an \lstinline{ExecutableExpression} object. Crucially, this class does not have any special operator or mathematical function overloads, which stops it from combining with other expressions, effectively ``freezing'' the expression.

As part of the transformation \lstinline{Eval()} also attempts to optimize the expression. It does so by performing common sub-expression elimination, constant folding, or replacing known patterns with more specialized equivalents.

The constructed \lstinline{ExecutableExpression} can then be executed via its \lstinline{Execute()} member function. This routine will perform any necessary pre- and post-processing (for example moving data to the GPU and calling \lstinline{PetscDeviceContextMarkIntentFromIDBegin/End()}) before finally launching the actual kernel. The manual construction and execution process is shown in \cref{code:managed:manual_construction}. 
\begin{lstlisting}[
  caption=Manual Expression Construction And Execution Example,
  label=code:managed:manual_construction
]
Petsc::ManagedScalar x, y, z, w;

// The type of expr1 encodes the 
// operation (x + y) / z, but expr
// itself is still symbolic
auto expr1 = (x + y) / z;
// It is possible to compose symbolic 
// expressions
auto expr2 = std::sin(expr1 * z) + 15;
// Evaluates the combined expression, 
// returning an ExecutableExpression object.
// Note the expanded form of expr2 is
// std::sin(((x + y) / z) * z) + 15
// which Eval() may simplify to
// std::sin(x + y) + 15
auto evaluated = Petsc::Eval(expr2);
// Execute the combined expression, storing 
// the result in w
evaluated.Execute(w);
// The expression may be executed any number
// of times, with any number of targets
evaluated.Execute(x);
\end{lstlisting}

\subsubsection{Solving The Scalar Problem}
\label{sec:design:managed:scalar_problem}
\noindent
Most users will not need to interface with the \lstinline{ExecutableExpression} directly. Instead, they are expected to use \lstinline{Eval()} in combination with the \lstinline{ManagedMemory} copy-constructor or copy-assignment operator, which will automatically make the necessary calls to evaluate the expression. This feature allows a \lstinline{ManagedMemory} object to be used as a nearly seamless drop-in replacement for scalars and scalar expressions, effectively solving the scalar problem.

A secondary goal of \lstinline{Eval()} (and for which it has a special overload) is to associate a \lstinline{PetscDeviceContext} with the construction of the \lstinline{ExecutableExpression}. This ensures that the expression is executed on the specified stream. This is required in the aforementioned \lstinline{ManagedMemory} constructor/assignment operator cases, where it is impossible or unwieldy to pass the \lstinline{PetscDeviceContext} as an additional argument.

Finally, as a convenience, when using the \lstinline{ManagedMemory} copy-constructor or assignment operator, the use of \lstinline{Eval()} may be omitted entirely. This is equivalent to using a globally blocking \lstinline{PetscDeviceContext}, which, of course, is implicitly synchronous. This is illustrated in \cref{code:managed:eval_optional}.
\begin{lstlisting}[
  caption=Example Showing Optional Eval Usage,
  label=code:managed:eval_optional
]
Petsc::ManagedReal x, y, z;
PetscDeviceContext dctx;

PetscDeviceContextCreate(&dctx);
PetscDeviceContextSetStreamType(dctx, PETSC_STREAM_GLOBAL_BLOCKING);

// The following are all equivalent to one
// another
x = Petsc::Eval(y + z, dctx);
x = Petsc::Eval(y + z, nullptr);
x = Petsc::Eval(y + z);
x = y + z;

// The following are all equivalent to one
// another
Petsc::ManagedReal w{y + z, dctx};
Petsc::ManagedReal w{y + z, nullptr};
Petsc::ManagedReal w{y + z};
\end{lstlisting}
An important point to note is the lifetime of a \lstinline{ManagedMemory}. Given their asynchronous nature, these objects may go out of scope before their various operations even occur, however, the underlying memory owned by the \lstinline{ManagedMemory} is kept alive until the corresponding stream idle. In other words, it is perfectly to safe for the user to destroy these objects without synchronizing, any remaining resources will be released once the stream is idle.
\subsection{Putting It All Together}
\label{sec:design:together}
\noindent
Equipped with \lstinline{PetscDeviceContext} and \lstinline{ManagedMemory} we are finally ready to tackle the integration problem writ large. To illustrate this, we convert the normalization code shown in \cref{code:vec_normalize}  to use the unified asynchronous framework, shown in \cref{code:managed:normalize_async}.
\begin{lstlisting}[
  caption=ManagedMemory Normalization of a Vector,
  label=code:managed:normalize_async
]
Petsc::ManagedReal alpha;
PetscDeviceContext dctx;

// Create a PetscDeviceContext, by default will use
// default blocking stream
PetscDeviceContextCreate(&dctx);
// Asynchronously compute the norm on the GPU, storing
// its future in alpha. dctx is used as the stream
VecNormAsync(v, NORM_2, &alpha, dctx);
// Evaluate the reciprocal in a kernel on device
alpha = Petsc::Eval(1.0 / alpha, dctx);
// Pass the updated future on to the next 
// asynchronous function
VecScaleAsync(v, alpha, dctx);
\end{lstlisting}
In addition to the use of the aforementioned objects, there is another key difference between \cref{code:vec_normalize} and \cref{code:managed:normalize_async}. The explicitly asynchronous \lstinline{Vec} API.

The routines \lstinline{VecNormAsync()} and \lstinline{VecScaleAsync()} shown in \cref{code:managed:normalize_async} are examples of stream-augmented versions of their regular counterparts. In place of \lstinline{PetscScalar} or \lstinline{PetscReal} arguments\footnote{Depending on PETSc configuration, \lstinline{PetscScalar} could be complex  or real, but \lstinline{PetscReal} is always real.}, 
these routines accept equivalently qualified \lstinline{ManagedScalar} or \lstinline{ManagedReal}s. They also accept a \lstinline{PetscDeviceContext} as the final argument, which dictates the stream to use for the operation. At a high level, the model can be summarized follows:
\begin{itemize}
    \item The user assigns high level operations to a stream by passing a \lstinline{PetscDeviceContext} as an argument to that operation.
    \item The underlying implementation uses the \lstinline{PetscDeviceContext} to mark its memory access patterns.
    \item The user can perform abitrary, asynchronous scalar arithmetic using the \lstinline{ManagedMemory} class.
    \item The user does not need to perform any stream dependency analysis. The marking system will automatically deduce any required serialization from the memory dependency graph, and transparently serialize streams.
    \item The user does not need to perform any explicit stream synchronization. Standard accessors (e.g. \lstinline{VecGetArray()}, \lstinline{VecGetArrayRead()}, \lstinline{MatDenseGetArrayWrite()}, etc.) will query the memory dependence graph, and implicitly perform synchronization if necessary.
\end{itemize}
The final two points have important consequences. First, they obviate the need to expose the internally recorded events to the user, since the user has no need to perform serialization or synchronization themselves. Second, they allow users to freely intermix asynchronous and synchronous API. The benefit is twofold. Not only does this allow incrementally porting synchronous to asynchronous, but it also means that asynchronous code is user-safe. So long as the downstream user goes through proper access channels and/or uses only high level functions, then it does not matter what asynchronous code precedes it. They cannot induce a race condition.

An example of is shown in \cref{code:sync_async_mix}, which illustrates how one might implement the main loop for a preconditioned Conjugate Gradient solver. We use both different streams and intermix synchronous and asynchronous calls to illustrate the simplicity of the programming model.
\begin{lstlisting}[
  caption=Example Showcasing Mixed API Usage To Implement The Main Loop For A Preconditioned Conjugate Gradient Solver,
  label=code:sync_async_mix
]
// initialization not shown for brevity
PetscDeviceContext   dctx_a, dctx_b, dctx_c;
Petsc::ManagedScalar a, b, beta, betaold;
Petsc::ManagedReal   dp;
Vec                  P, Z, W, X, R;
Mat                  A;

for (PetscInt i = 1; i < max_it; ++i) {
  // Asynchronous, evaluation runs on GPU
  b = Petsc::Eval(beta / betaold, dctx_a);
  // Synchronous, b.front() returns a value on the 
  // host, so will perform implicit synchronization 
  // before returning value
  VecAYPX(P, b.front(), Z); // p <- z + b* p
  // Synchronous, may run on GPU or CPU, but since there 
  // is not PetscDeviceContext attached, will automatically
  // wait for asynchronous AYPX() to complete in either case
  MatMult(A, P, W); // w <- Ap
  // Asynchronous, back on GPU
  VecDotAsync(P, W, &a, dctx_b); // a <- p'w
  // Also asynchronous, will wait for Dot()
  a = Petsc::Eval(beta / a, dctx_c); // a <- beta / a
  // Asynchronous, on GPU, will execute concurrently
  // with expression above!
  betaold = Petsc::Eval(beta, dctx_a); // betaold <- beta
  // All 3 expressions are run on separate streams. Not only
  // will the negation of a wait for assignment of a above, 
  // but the AXPY will asynchronously wait for negation!
  VecAXPYAsync(X, a, P, dctx_b); // x <- x + ap
  VecAXPYAsync(R, -a, W, dctx_c); // r <- r - aw
  // Waits only on the second AXPY (as there is no 
  // dependence on X, a, or P!
  ApplyPreconditioner(R, Z); // z <- Br
  VecNormAsync(Z, NORM_2, &dp, dctx_a); // dp <- z'*z
  // User-supplied callback, do not need to synchronize
  (*user_converged_callback)(..., dp, &done);
  if (done) break; // converged
  VecDotAsync(Z, R, &beta, dctx_b); // beta <- z'*r
}
// No need to synchronize, OK for a, b, beta, and betaold
// to go out of scope. They will asynchronously release 
// their resources once the stream becomes idle!
\end{lstlisting}

\section{Experimental Evaluation}
\label{sec:results}
\noindent
To measure the performance of the asynchronous framework, we devise a series of benchmarks to study its behavior under various circumstances. Benchmarking is conducted using the Argonne Leadership Computing Facility's ``Polaris''. Each node of Polaris is equipped with 1 AMD EPYC ``Milan'' processor and 4 NVIDIA A100 GPUs, for a combined peak performance of 78 Teraflops in double precision. As AMD HIP has an almost identical programming model to CUDA we only show results using CUDA. We use GCC 11.2.0 as the host compiler, coupled with NVIDIA NVCC V11.4.100 (CUDA Toolkit 11.4) as the device compiler. Python benchmarks use JAX~\citep{jax2018github} version 0.4.12, together with CUDA Toolkit 11.7.

Our benchmarks are limited to linear solvers; not only do these represent the core of PETSc, they also present all of the necessary challenges to fully stress-test the framework.

\subsection{Methodology}
\label{sec:results:method}
\noindent
The benchmark considers the solution of a sparse, symmetric, and positive definite linear system arising from a finite-difference discretization of the usual constant coefficient Laplace equations. Here, we are not interested in the actual solution of the equations, but rather in the performance of the linear solver.

The problem is solved in both two and three dimensions, and with different stencil types: 5- and 9-points stencil in two dimensions, 7- or 27-point stencil in three dimensions. This allows us to investigate the effects of both work density and work size on runtime. To simplify comparison, each linear solve is run for 20 iterations, with Jacobi (diagonal scaling) preconditioning, and using only one MPI rank. For each operator configuration, we test three variants:
\begin{itemize}
    \item \verb|main_gpu_*|: The solver implementation on the current PETSc \verb|main| \verb|git| branch, run on the GPU. The baseline.
    \item \verb|main_cpu_*|: The solver implementation on the current PETSc \verb|main| \verb|git| branch, run on the CPU.
    \item \verb|async_*|: The re-implementation of the solver in the proposed asynchronous framework.
\end{itemize}
Each unique configuration constitutes one ``run''. In order to minimize interference, each run is performed as a separate invocation of a benchmarking script. This eliminates any possibility for resource caching, either on the GPU or within PETSc itself.

The benchmark consists of a single warm-up iteration -- so that one-time setup costs do not skew the results -- and a series of timing iterations. The actual timing is done by measuring the end-to-end duration of a call to \lstinline{KSPSolve()}. An approximate reconstruction of the timing loop is shown in \cref{code:timing_loop}.
\begin{lstlisting}[
  caption=Approximate Reconstruction of the Timing Loop,
  label=code:timing_loop
]
std::vector<PetscLogDouble> times;

for (PetscInt i = 0; i < nit; ++i) {
  PetscLogDouble begin, end;

  // Reset any operators (e.g. b, x) to 
  // original values, ensure device is idle,
  // and any cached values are invalidated
  PreSolve(ksp, b, x);
  PetscTime(&begin);
  // Do the solve
  KSPSolve(ksp, b, x);
  cudaDeviceSynchronize();
  PetscTime(&end);
  // Check correctness of solution (e.g.
  // compare residual norm to that of previous
  // iterations)
  PostSolve(ksp, b, x);
  times.emplace_back(end - begin);
}
\end{lstlisting}
Once the timing loop is finished, key metrics such as the total, average, maximum, and minimum times are computed. Among these, the \emph{minimum} time is used as the definitive time, as it is the most statistically stable. Additional statistics such as the number of \emph{floating point operations per second} (FLOPS), and the number of host-to-device/device-to-host copies (and their quantities) are collected after the program finished, by parsing PETSc's detailed logging output. Final post-processing of the results, such as computing the memory bandwidth rate, \emph{degrees of freedom} (DOF) solved per second, etc. is done \emph{en masse}, once all runs for a particular solve type were collected.
\subsection{Conjugate Gradient (CG)}
\label{sec:results:cg}
\begin{figure}[!ht]
    \centering
    \includegraphics[width=\linewidth,keepaspectratio]{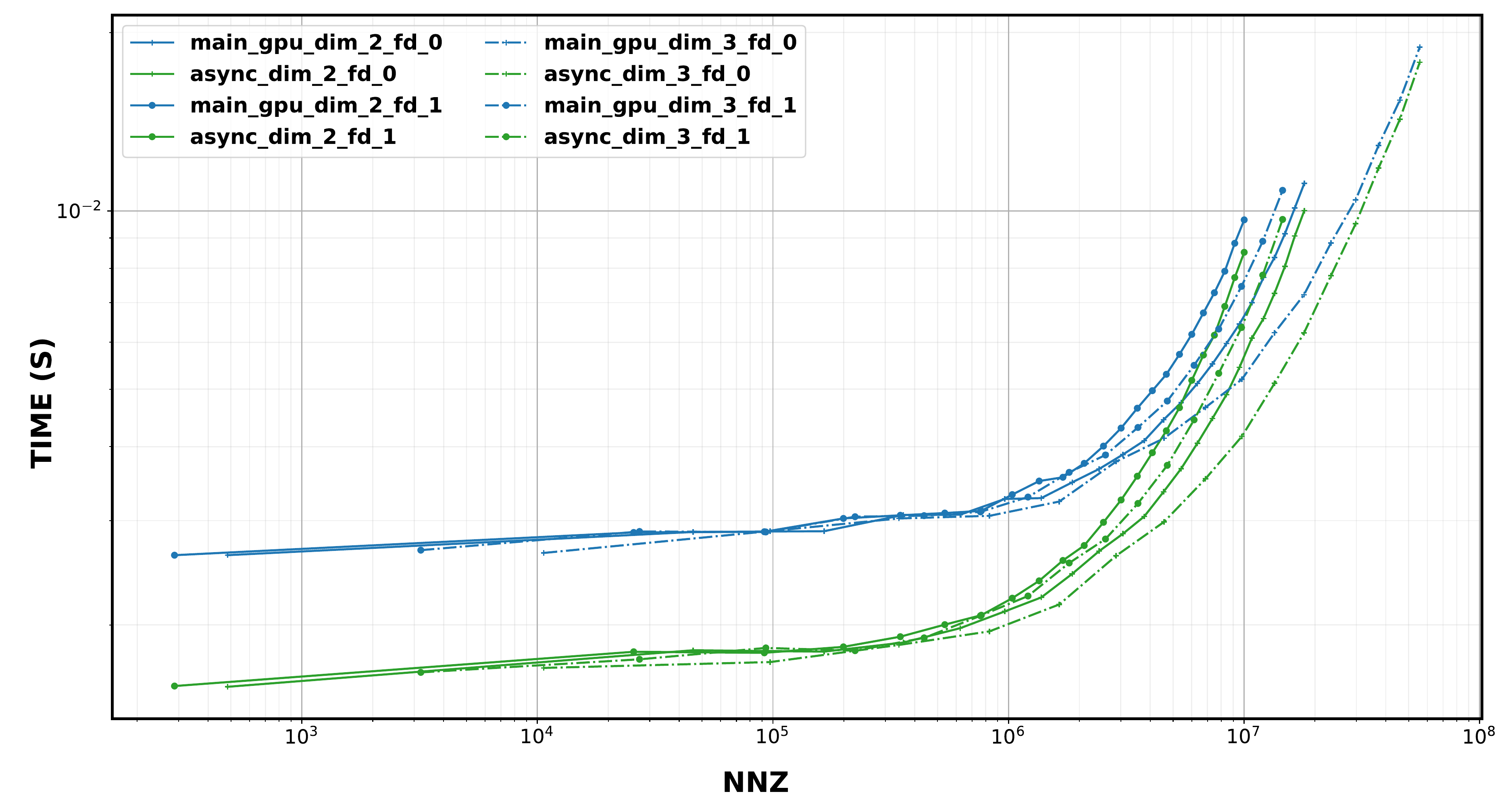}
    \caption{CG Time to solution (in seconds) versus NNZ. Lower is better.}
    \label{fig:cg:tts}
\end{figure}

\begin{figure}[!t]
    \centering
    \includegraphics[width=\linewidth,keepaspectratio]{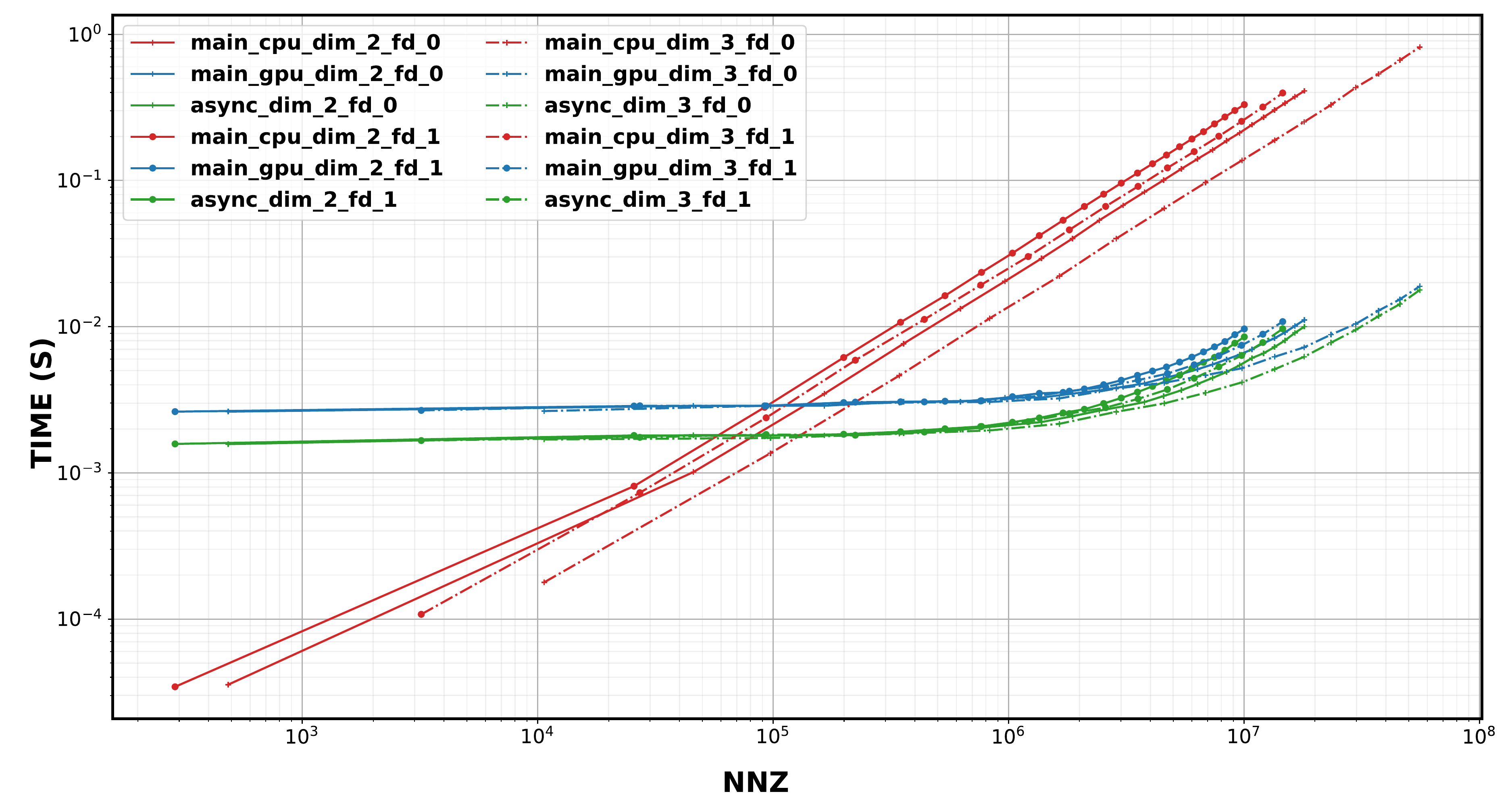}
    \caption{CG Time to solution (in seconds) versus NNZ, including CPU implementation. This illustrates that regimes where serialization overhead would be significant are dominated by CPU anyways. Lower is better.}
    \label{fig:cg:tts_cpu}
\end{figure}

\begin{figure}[!t]
    \centering
    \includegraphics[width=\linewidth,keepaspectratio]{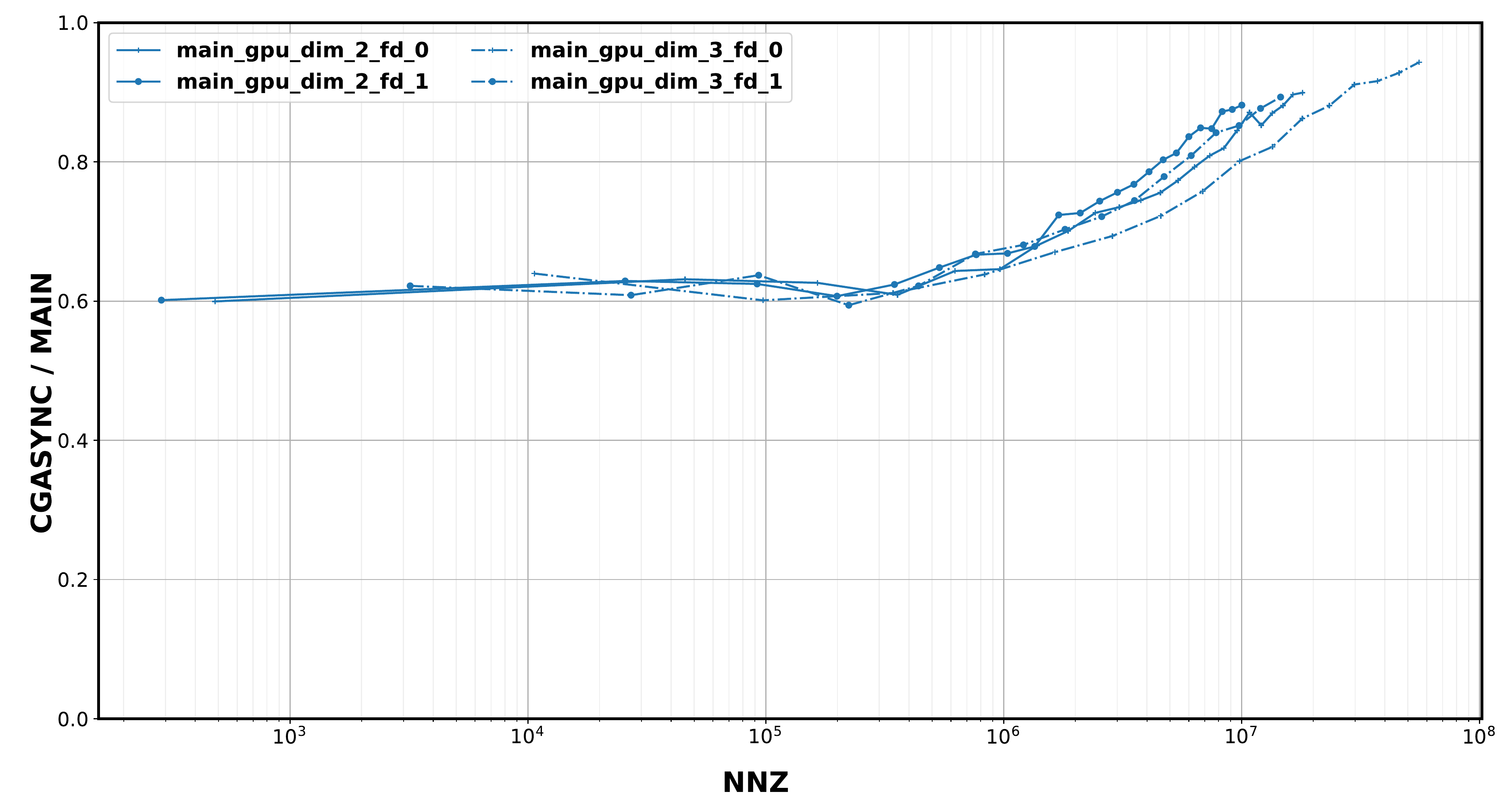}
    \caption{CG Ratio of time to solution versus NNZ. $y > 1$ indicates that async is slower than baseline, $y = 1$ indicates times are equal, and $y < 1$ indicates async is faster than baseline.}
    \label{fig:cg:ratio}
\end{figure}

\begin{figure}[!t]
    \centering
    \includegraphics[width=\linewidth,keepaspectratio]{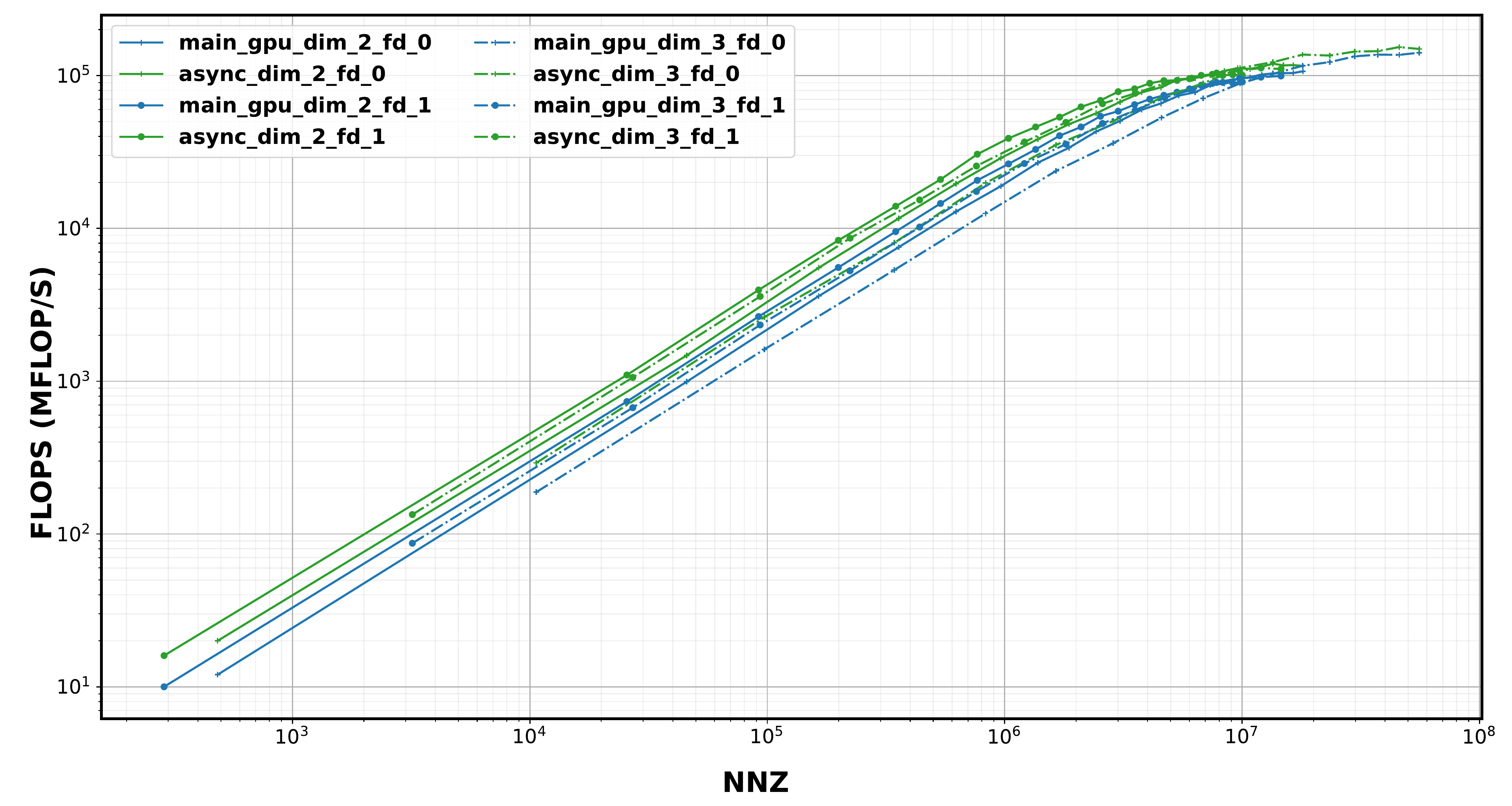}
    \caption{CG FLOPS versus NNZ. Higher is better.}
    \label{fig:cg:flops}
\end{figure}

\begin{figure}[!t]
    \centering
    \includegraphics[width=\linewidth,keepaspectratio]{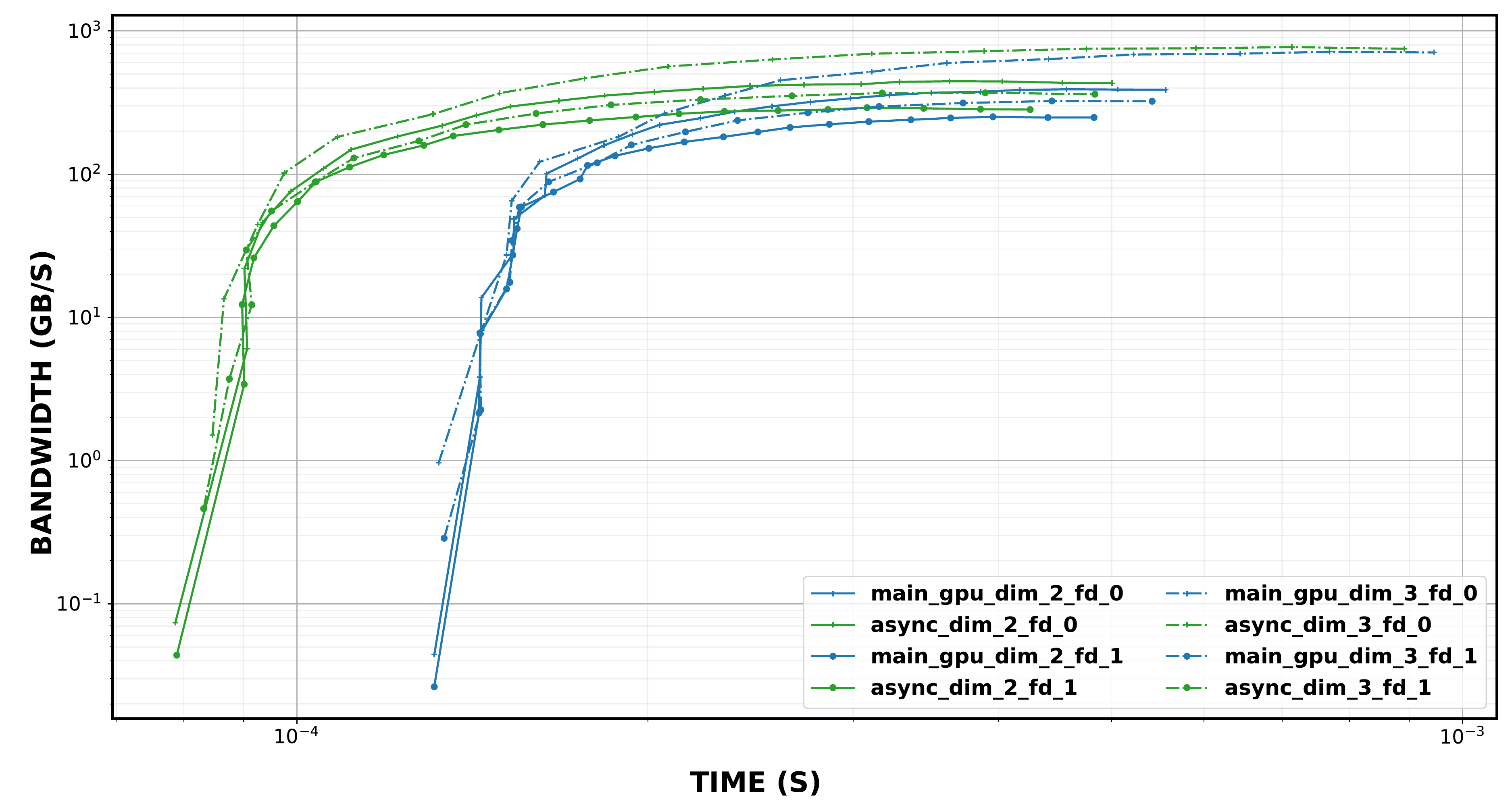}
    \caption{CG Memory bandwidth per iteration (in gigabytes per second) versus NNZ. Higher is better.}
    \label{fig:cg:bandwidth_cpu}
\end{figure}
\noindent
Our first benchmark examines the performance when using the conjugate gradient (CG) method~\citep{hestenes1952methods} as the linear solver.
CG is a type of Krylov subspace method (arguably the original such method for linear systems~\citep{saad2022krylovhistory}) and remains one of the most popular iterative algorithms for solving symmetric (or Hermitian) sparse linear systems.
Part of the popularity of CG (and similar methods) stems from its relative ease of implementation:
Each iteration of CG requires only one matrix-vector product, three inner products, some scalar operations, and three vector updates. 

\Crefrange{fig:cg:tts}{fig:cg:bandwidth_cpu} detail the results collected. The cost of performing matrix-vector products is proportional to the number of nonzero (NNZ) entries in the matrix, so it is a common metric for estimating the amount of computational work required per iteration. \Cref{fig:cg:tts} shows the solve time versus NNZ. The async implementation is faster than the baseline implementation at every measured point.

\Cref{fig:cg:tts_cpu} shows the same solve time versus NNZ also including the times for the CPU implementation, which is faster than both the baseline and asynchronous variant for NNZ less than $\approx 10^5$. It serves to demonstrate that any moderate overhead in dynamically determining memory dependencies ultimately does not matter. Such overhead is only relevant when GPU kernels are so short that they finish before the CPU has time to queue up the next one. In our case, this corresponds to instances where the matrix is very small or very sparse. Since the CPU implementation is faster in this regime, the overhead does not play a role, since the GPU implementation would not be used anyways.

\Cref{fig:cg:ratio} depicts the ratio between the async and baseline times. The async variant achieves a maximum speedup of 1.8x over the baseline implementation. However, this speedup diminishes when NNZ is larger than  $\approx 10^6$; in this regime, the cost of the matrix-vector multiplication starts to dominate the overall runtime. 

\begin{figure}[!t]
    \centering
    \includegraphics[width=\linewidth,keepaspectratio]{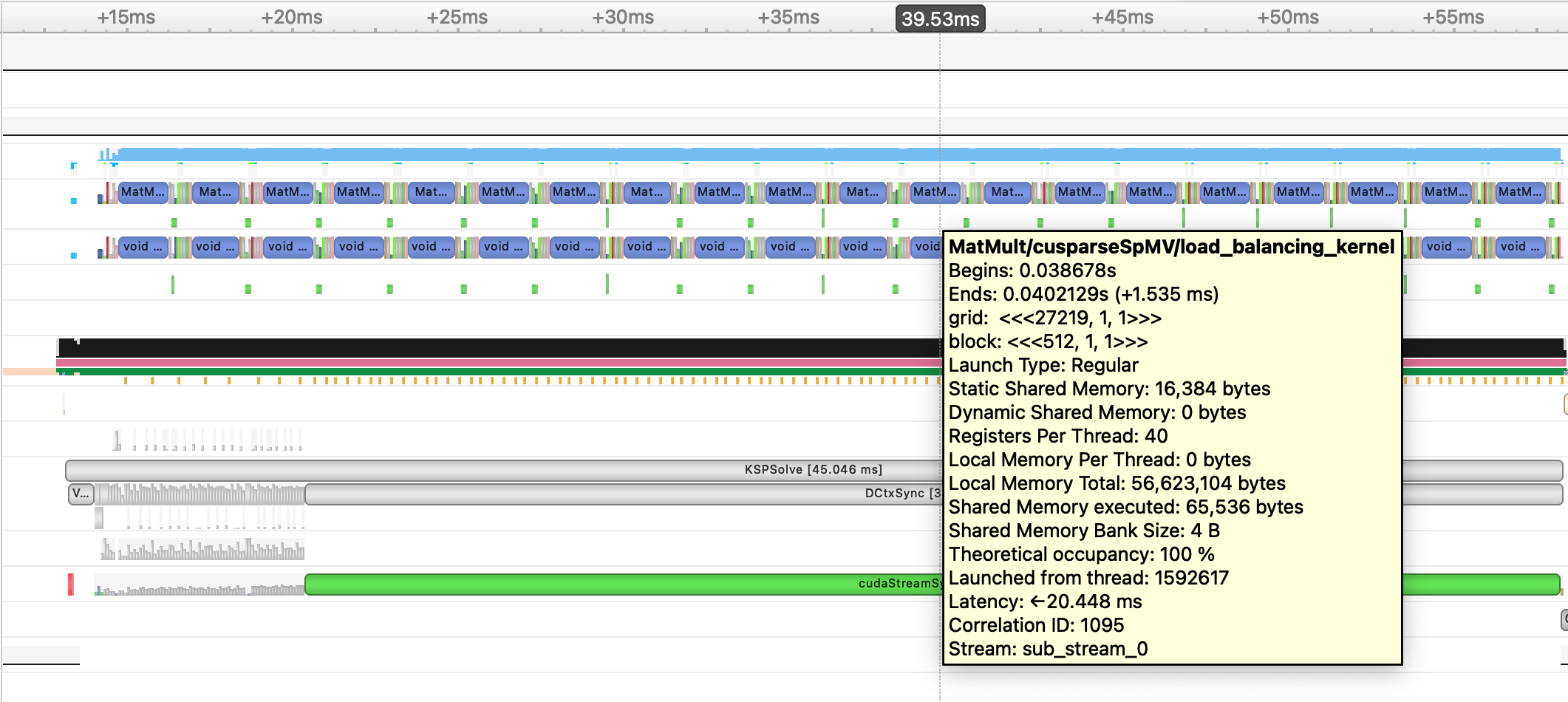}
    \caption{Timeline view of a CG solve for NNZ $\geq 10^6$ to illustrate that the time for matrix-vector multiplication (shown as blue blocks) dominates the overall solve time for large matrices.}
    \label{fig:cg:timeline}
\end{figure}

This is illustrated in \Cref{fig:cg:timeline}, which shows a timeline view of a particular solve. Each GPU kernel is depicted as a solid block along the first and second rows in the timeline. The size of each block is proportional to the runtime of the kernel. The CPU functions are depicted on the ninth row as a series of grey blocks. As with the GPU kernels, the size of each block denotes its runtime.

The matrix-vector product kernels are colored in deep blue, and are proportionally much larger than any other kernel block in the timeline, indicating that they make up the vast majority of GPU runtime.

\subsection{Transpose-Free Quasi-Minimal Residual (TFQMR)}
\label{sec:results:tfqmr}
\begin{figure}[!t]
    \centering
    \includegraphics[width=\linewidth,keepaspectratio]{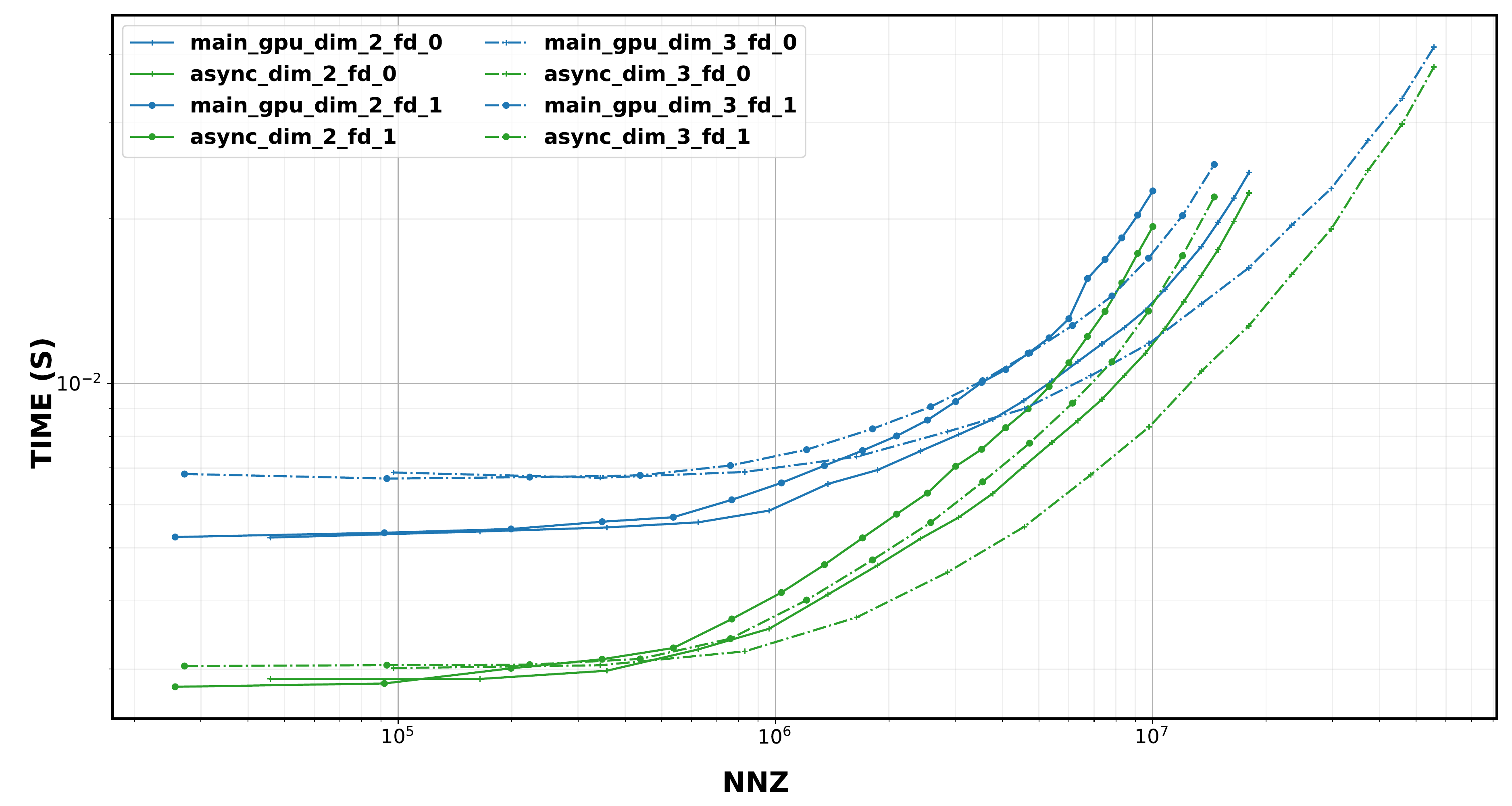}
    \caption{TFQMR Time to solution (in seconds) versus NNZ. Lower is better.}
    \label{fig:tfqmr:tts}
\end{figure}

\begin{figure}[!t]
    \centering
    \includegraphics[width=\linewidth,keepaspectratio]{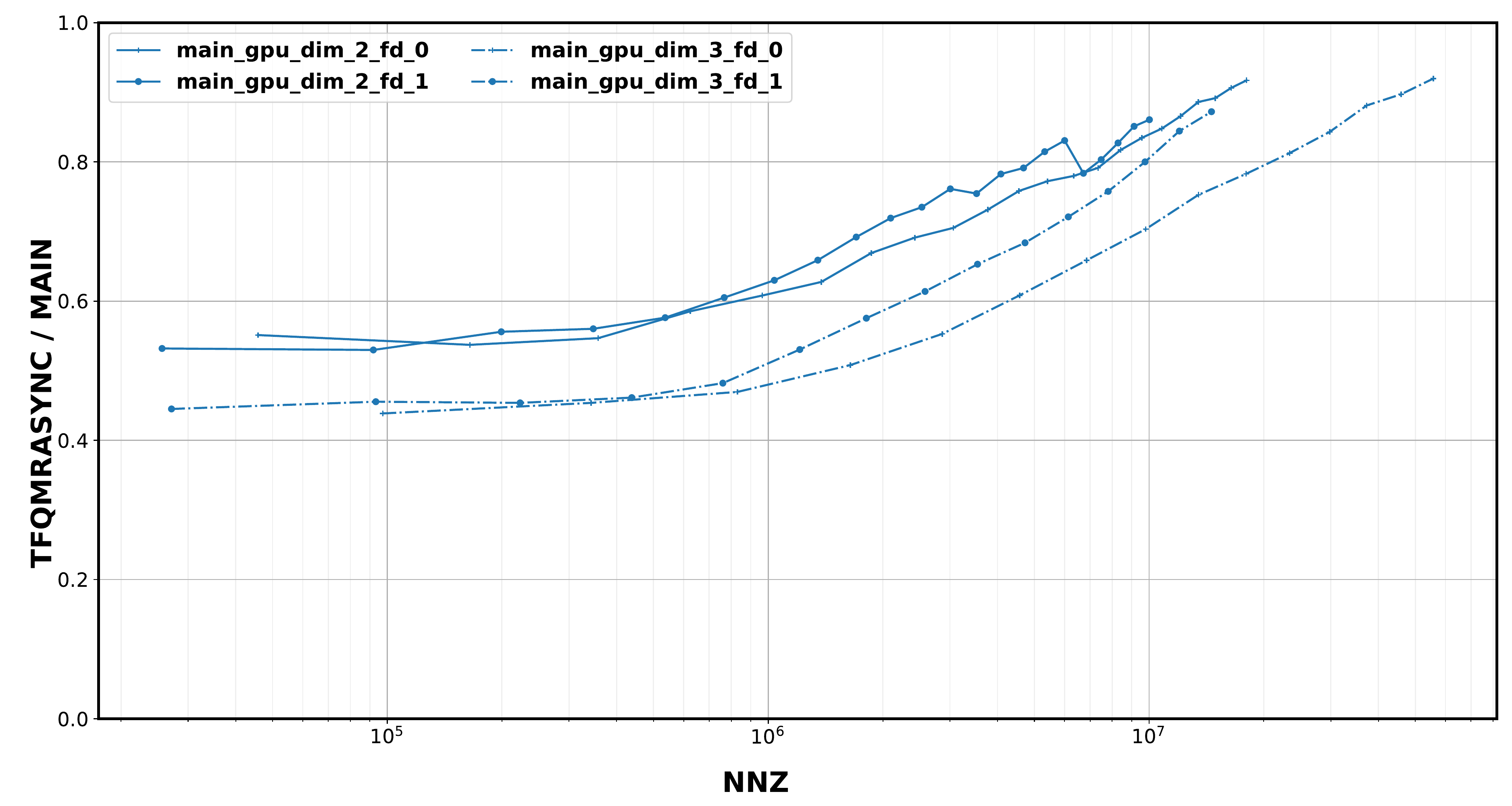}
    \caption{TFQMR Ratio of time to solution versus NNZ. $y > 1$ indicates that async is slower than baseline, $y = 1$ indicates times are equal, and $y < 1$ indicates async is faster than baseline.}
    \label{fig:tfqmr:ratio}
\end{figure}

\begin{figure}[!t]
    \centering
    \includegraphics[width=\linewidth,keepaspectratio]{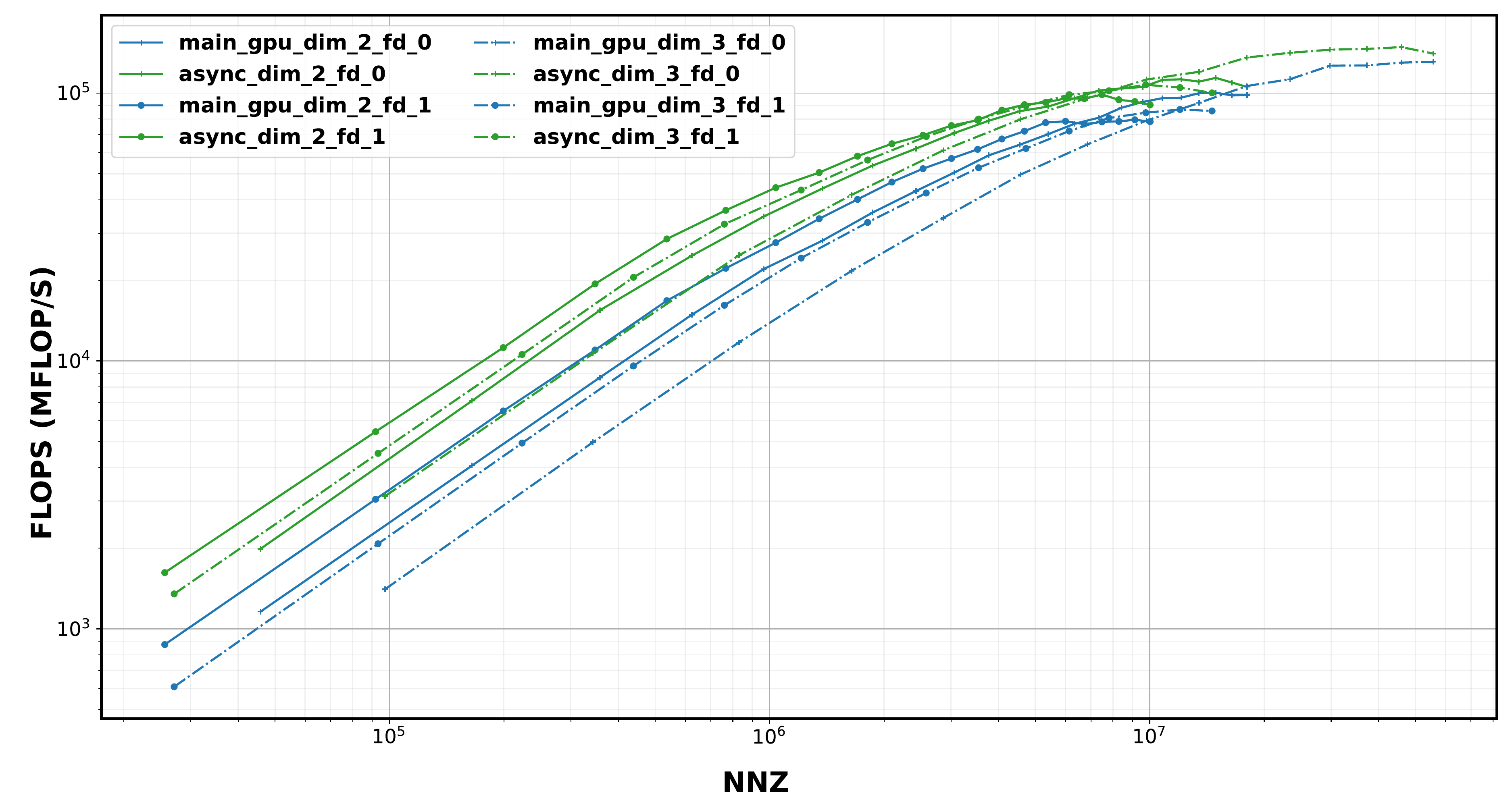}
    \caption{TFQMR FLOPS versus NNZ. Higher is better.}
    \label{fig:tfqmr:flops}
\end{figure}

\begin{figure}[!t]
    \centering
    \includegraphics[width=\linewidth,keepaspectratio]{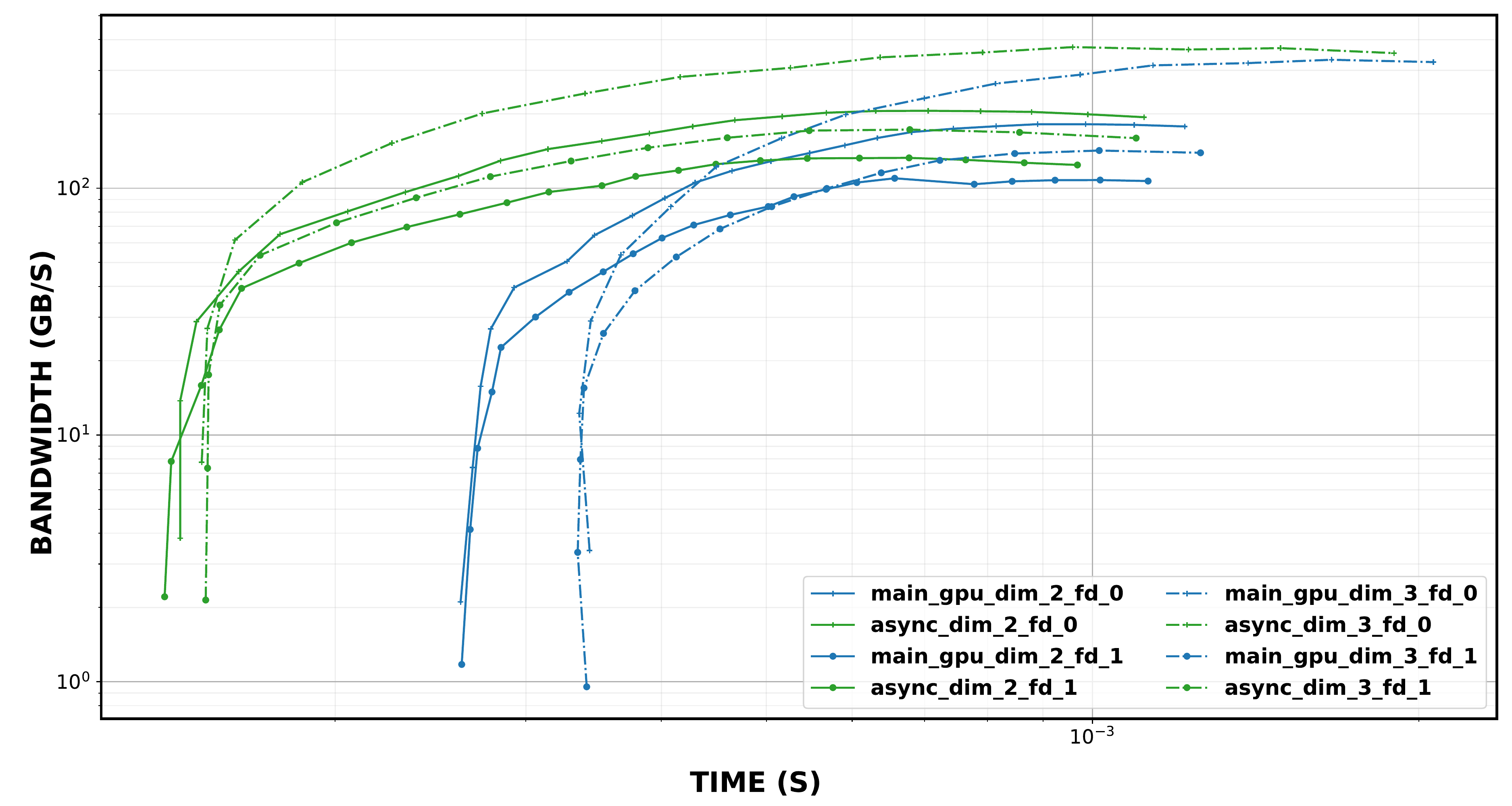}
    \caption{TFQMR Memory bandwidth per iteration (in gigabytes per second) versus NNZ. Higher is better.}
    \label{fig:tfqmr:bandwidth_cpu}
\end{figure}
\noindent
The second performance benchmark we consider solves the same Laplace problem solved in the first benchmark, but using the transpose-free quasi-minimal residual method (TFQMR) \citep{freund1993tfqmr} that, unlike CG, is applicable to indefinite and non-Hermitian systems.
In contrast to the popular GMRES algorithm \citep{saad-schultz1986gmres}, many other Krylov methods are based on a short-term recurrence, which avoids explicitly building a set of basis vectors for orthogonalization of the Krylov subspace and does not require an upper Hessenberg matrix solve.
This simplicity makes TFQMR, and other non-GMRES-like Krylov methods, a popular choice for non-Hermitian solvers on GPUs \citep{adams2023performance,kashi_nayak_kulkarni_scheinberg_lin_anzt_2022}.
In terms of computational kernels required, a TFQMR iteration is similar to a CG iteration, except that two matrix-vector products and roughly twice the number of vector inner products are performed.

\Crefrange{fig:tfqmr:tts}{fig:tfqmr:bandwidth_cpu} depict performance results of the asynchronous TFQMR implementation versus the baseline. As with CG, the async implementation is faster than the baseline at every measured point. In fact, it shows an even greater improvement than CG; for smaller problems (NNZ $\lessapprox 10^6$) it is nearly twice as fast (\Cref{fig:tfqmr:ratio}) versus only an 80\% improvement in CG (\Cref{fig:cg:ratio}). Just like CG, this improvement tapers off as the matrices became larger, for mostly the same reason -- the cost of performing the matrix-vector product.
\subsection{Comparisons Against JAX}
\label{sec:results:jax}
\begin{figure}[!t]
    \centering
    \includegraphics[width=\linewidth,keepaspectratio]{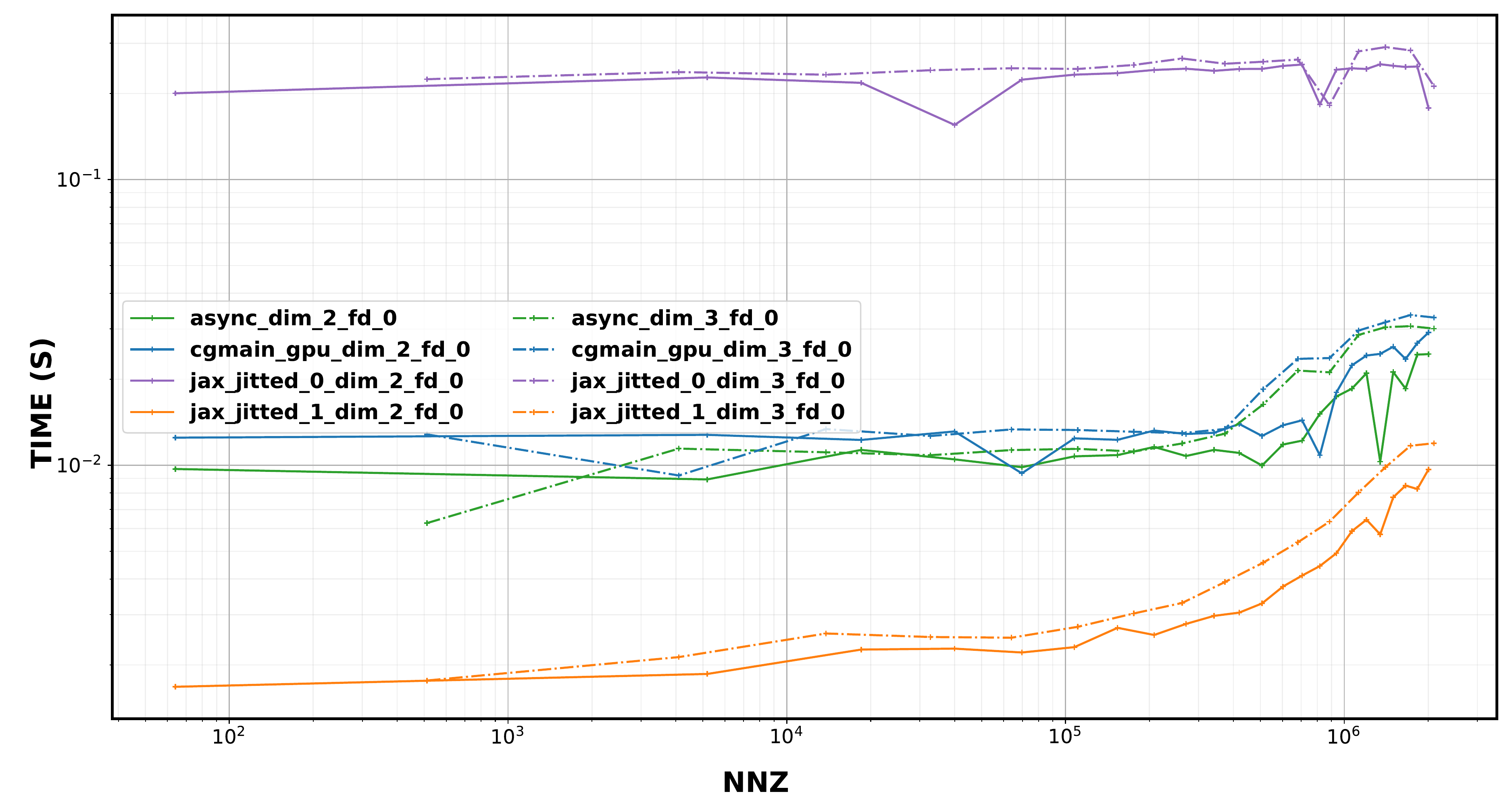}
    \caption{Comparing CG time to solution (in seconds) versus NNZ against JAX. Lower is better.}
    \label{fig:jax:tts}
\end{figure}
\noindent
Finally, we also compared the CG solver against a popular Python-based Just-In-Time (JIT) library, JAX, to determine a practical upper performance bound. In this benchmark, we use the Python bindings of PETSc and perform the matrix-vector multiplication matrix-free in Python code by using convolution primitives from JAX.

Results are shown in \Cref{fig:jax:tts}: {\bf jax\_jitted\_1} correspond to the timings associated with JIT-ing 20 iterations of CG into a single call, while {\bf jax\_jitted\_0} indicates that only the matrix-vector product and the Jacobi scaling are JIT-ed, while the asynchronous CG procedure in JAX is run in the Python interpreter. The PETSc async implementation runs CG from the PETSc library and uses the JIT-ed version of the matrix-vector product, while Jacobi scaling is performed using PETSc library calls.  

The async code is an order of magnitude faster than {\bf jax\_jitted\_0}, while it is slower -- as expected -- than the full JIT-ed version  {\bf jax\_jitted\_1}. In this case, the JIT'ed version is over 5x faster than our implementation for small problem sizes. It is interesting to note, however, that as the problem size increases the advantage of full JIT-ing diminishes. JAX's speedup steadily drops to $\sim$2x over our implementation for NNZ $\gtrapprox 10^6$. 

These comparisons illustrate that our programming model is able to get reasonably close to a fully JIT'ed code while retaining the full flexibility of a composed version. PETSc timings are less regular than those shown in \Cref{sec:results:jax} since JAX uses its own streams internally, and we synchronize the matrix-vector stream every time before copying back the result of the operation into PETSc-managed vectors. Better handling of the PETSc-JAX interface may slightly improve the results shown, particularly for smaller problem sizes.
\section{Conclusion And Future Work}
\label{sec:conclusion}
\noindent
This work develops a library programming model that exposes streams to the algorithmic design. The design of a programming model, like a language, is a complex endeavor with many competing criterion, and the design presented here will evolve as we gain experience in its application. Nevertheless, this work demonstrates its effectiveness and scalability, and shows that it provides broad performance benefits. 

We introduced two new library objects -- \lstinline{PetscDeviceContext}, and \lstinline{ManagedMemory} -- described their implementation, API, and use, in a generic asynchronous programming model. Our experiments established the efficiency of this programming model and that of the newly introduced objects in the context of a general-purpose library like PETSc.

Future work shall focus on multi-GPU settings and incorporate MPI into the asynchronous framework. As discussed in \cref{sec:problems:async_problem}, the current GPU-aware MPI specification is incapable of handling data produced on a user-driven device stream. At present, the user must synchronize their stream before making collective MPI calls. This remains a major hurdle in achieving parallel performance.

Another major problem for asynchronous frameworks is related to dynamic stopping criteria. It is not presently possible to ``unlaunch'' a kernel once it has been queued, and so it is difficult to implement, e.g., asynchronous convergence checks. Future work will focus on possible solutions to this problem, such as semaphores that subsequent kernels check before executing. This will lay the groundwork for fully asynchronous nonlinear solvers and ordinary differential equations integrators.

\section*{Acknowledgments}
\addcontentsline{toc}{section}{Acknowledgments}
\noindent
This research was supported by the Exascale Computing Project (17-SC-20-SC), a collaborative effort of the U.S. Department of Energy Office of Science and the National Nuclear Security Administration. This research used resources of the Argonne Leadership Computing Facility, which is a DOE Office of Science User Facility supported under Contract DE-AC02-06CH11357. This research was supported by the Office of Advanced Scientific Computing Research, Scientific Discovery through Advanced Computing (SciDAC) program through the FASTMath Institute under Contract No. DE-AC02-05CH11231 at Lawrence Berkeley National Laboratory.

\section*{Code Availability}
\noindent
Code related to this work may be found at \url{https://gitlab.com/petsc/petsc}. Data was collected using the \verb|main|, and \verb|jacobf/2022-11-28/petsc-managed-memory| \verb|git| branches respectively. Benchmarking, post-processing, and plotting was performed using scripts found at \url{https://gitlab.com/Jfaibussowitsch/async_petsc}.

\bibliographystyle{elsarticle-num}
\bibliography{references}

\parbox[t]{\textwidth}{
\begin{wrapfigure}{l}{25mm}
  \includegraphics[width=1in,height=1.25in,clip,keepaspectratio]{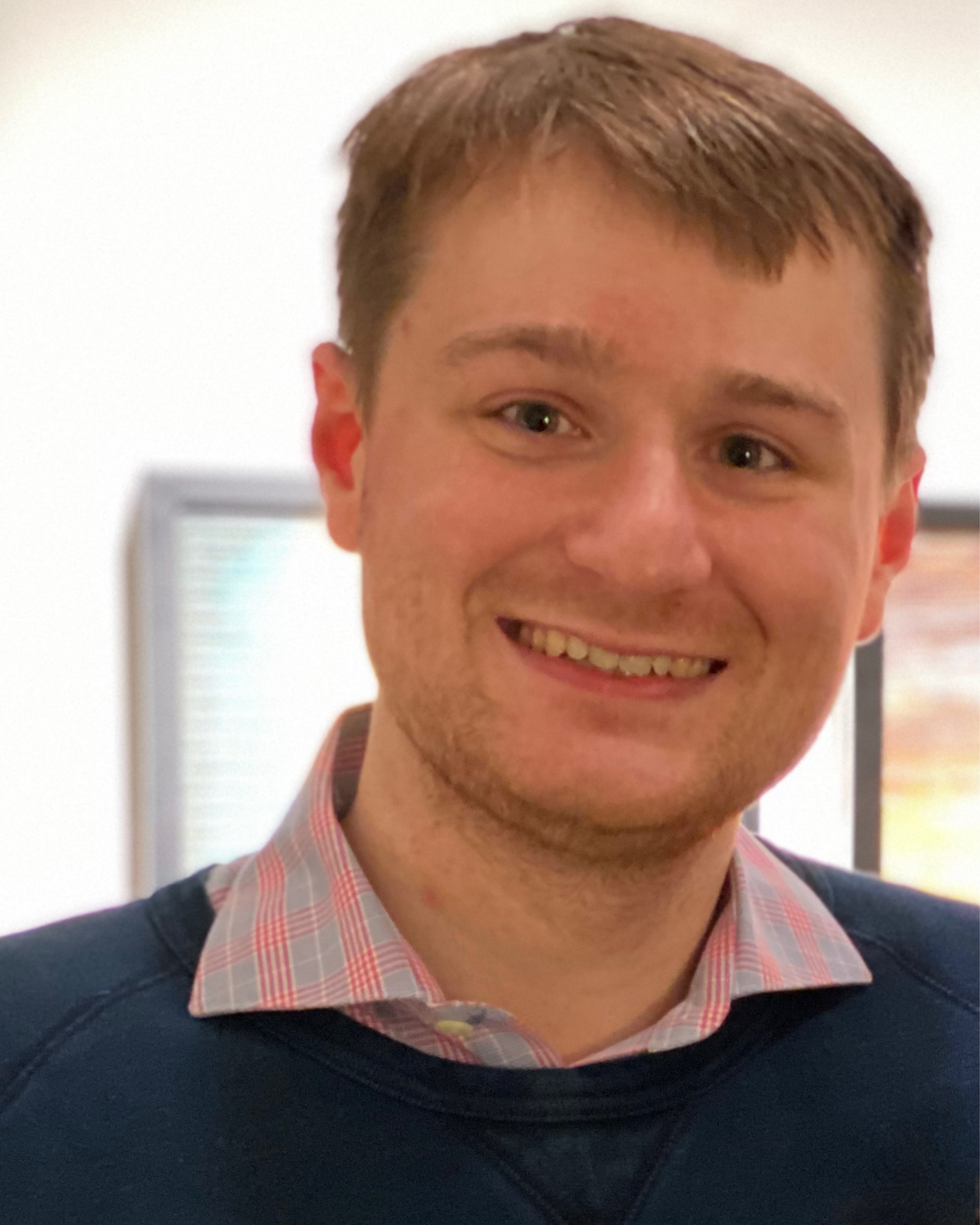}
\end{wrapfigure}
\noindent\singlespacing
\textbf{Jacob~Faibussowitsch} is a Research Software Engineer at Argonne National Laboratory and a PETSc developer. His research focuses on computational software engineering, in particular the development of novel, high-performance GPU frameworks. He earned his M.S. in Mechanical Engineering at the University of Illinois at Urbana-Champaign.}
\vspace{0.1\baselineskip}

\parbox[t]{\linewidth}{
\begin{wrapfigure}{l}{25mm} 
\includegraphics[width=1in,height=1.25in,clip]{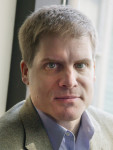}
\end{wrapfigure}
\noindent\singlespacing
\textbf{Mark~F.~Adams} received his Ph.D. in Civil Engineering, from U.C. Berkeley and is currently a Staff Scientist in the Scalable Solvers Group at Lawrence Berkeley National Laboratory.}
\vspace{\baselineskip}

\parbox[t]{\linewidth}{
\begin{wrapfigure}{l}{25mm} 
\includegraphics[width=1in,height=1.25in,clip,keepaspectratio]{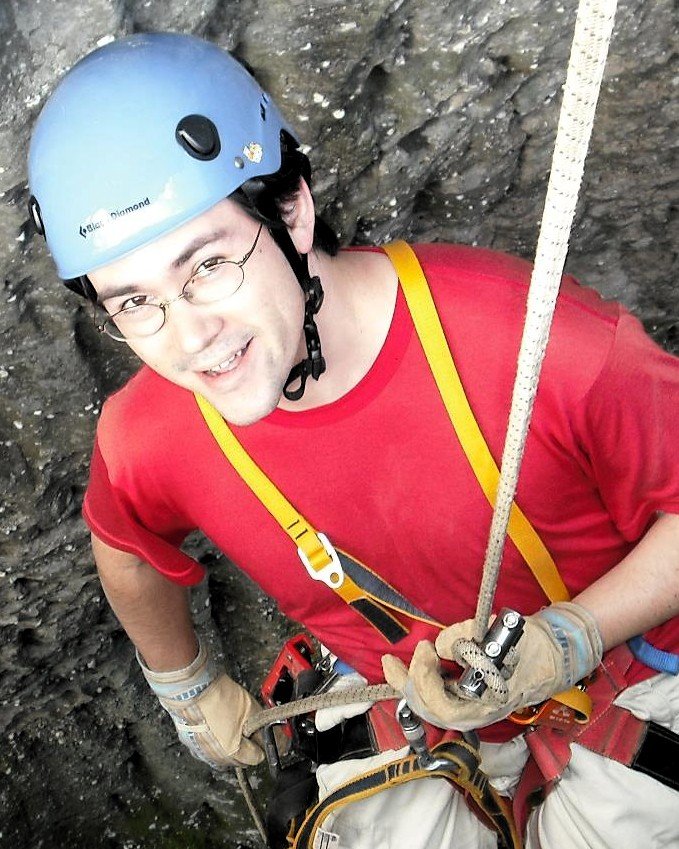}
\end{wrapfigure}
\noindent\singlespacing
\textbf{Richard~Tran~Mills} is a computational scientist at Argonne National Laboratory. His research spans high-performance scientific computing, machine learning, and the geosciences. He is a developer of PETSc and the hydrology code PFLOTRAN. He earned his Ph.D. in computer science at the College of William and Mary.}
\vspace{\baselineskip}

\parbox[t]{\linewidth}{
\begin{wrapfigure}{l}{25mm}
\includegraphics[width=1in,height=1.25in,clip,keepaspectratio]{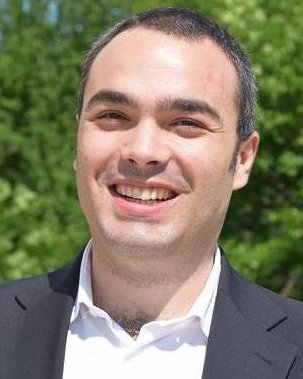}
\end{wrapfigure}
\noindent\singlespacing
\textbf{Stefano~Zampini} is a research scientist in the Extreme Computing Research Center of King Abdullah University for Science and Technology (KAUST), Saudi Arabia. He received his Ph.D. in applied mathematics from the University of Milano Statale. He is a developer of PETSc.}
\vspace{\baselineskip}

\parbox[t]{\linewidth}{
\begin{wrapfigure}{l}{25mm} 
\includegraphics[width=1in,height=1.25in,clip,keepaspectratio]{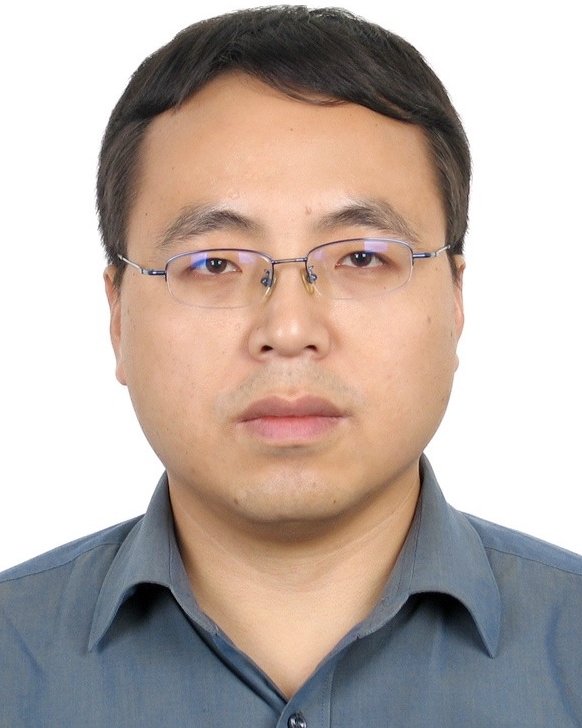}
\end{wrapfigure}
\noindent\singlespacing
\textbf{Junchao~Zhang} is a Research Software Engineer at Argonne National Laboratory. He is a PETSc developer and works mainly on
communication and GPU support in PETSc. He received his Ph.D. in computer science from the Chinese Academy of Sciences, Beijing, China.}

\end{document}